\documentclass[a4paper,11pt]{article}
\pdfoutput=1
\usepackage[T1]{fontenc}
\usepackage{array}
\usepackage{amsbsy}
\usepackage{amstext}
\usepackage{amsmath}
\usepackage{amssymb}
\usepackage{graphicx}
\usepackage{color}

\newcommand\ce{\,\!}

\newcommand{\negminispace}{\kern-.016667em} 

\newcommand{\half}{\kern.083333em}   
\newcommand{\quart}{\kern.0416675em}  
\newcommand{\nhalf}{\kern-.083333em}   
\newcommand{\nquart}{\kern-.0416675em}  



\newcommand{\smallcaption}[1]{\caption{\protect\small#1}}

\newcommand\apar{a_\parallel}
\newcommand\aperp{a_\perp}

\renewcommand{\d}{\text{d}}

\newcommand\curl{\mathop{\rm curl}\nolimits}
\newcommand\curlop[1]{\curl\nhalf #1} 
\newcommand\curlH{\curlop{H}}

\begin{document}

\title{Analytic Analysis of Irregular Discrete Universes}

\author{Shan W. Jolin \\ \emph{Nanostructure Physics, Royal Institute of Technology} \\ \emph{106 91 Stockholm, Sweden} \and  Kjell Rosquist \\ \emph{Department of Physics, Stockholm University} \\  \emph{106 91 Stockholm, Sweden}} 
%
%

\date{}
\maketitle

\begin{center}
\begin{minipage}[t]{0.8\linewidth}\small{\centering
\emph{Abstract}\\[5pt]
In this work we investigate the dynamics of cosmological models with spherical topology containing up to 600 Schwarzschild black holes arranged in an irregular manner.
We solve the field equations by tessellating the 3-sphere into eight identical cells, each having a single edge which is shared by all cells.  
The  shared edge is enforced to be locally rotationally symmetric (LRS), thereby allowing for solving the dynamics to high accuracy along this edge.
Each cell will then carry an identical (up to parity) configuration which can however have an arbitrarily random distribution. The dynamics of such models is compared to that of previous works on regularly distributed black holes as well as with the standard isotropic dust models of the FLRW type. The irregular models are shown to have richer dynamics than that of the regular models. The randomization of the distribution of the black holes is done both without bias and also with a certain clustering bias. The geometry of the initial configuration of our models is shown to be qualitatively different from the regular case in the way it approaches the isotropic model.}
\end{minipage}
\end{center}



{\vspace{1.5cm}

\section{Introduction}

matter distribution in the universe is discrete and therefore strongly locally inhomogeneous, it is natural to ask how this affects the overall expansion rate. The question is to what extent the dynamics of the real universe is described by the homogeneous fluid FLRW models. Due to the nonlinear nature of the Einstein equations this is a difficult problem often referred to as the averaging problem in general relativity \cite{Ellis_Buchert,wiltshire}. It has been extensively discussed in the past but there is still no consensus about its solution (see e.g. \cite{Buchert_etal}). A new avenue to address this issue was recently opened by considering a universe filled with Schwarzschild black holes as sources \cite{Clifton_etal:2012,Bentivegna,Clifton_etal:2013,Clifton_etal:2017,Bentivegna 2,BHL_review}. The idea is to first specify the sources as initial data on a time-symmetric spatial hypersurface with spherical topology. It is possible to do this using a finite number of Schwarzschild black holes as an exact solution of the Einstein constraints. This configuration of initial data can then in principle be evolved using the Einstein evolution equations. While the general evolution can only be done numerically, some subsets with sufficiently high symmetry can be evolved exactly or approximately by analytical methods. The line of investigation is still in its infancy with no clearcut answer to the averaging issue. However, we believe that this approach has a promising potential in this respect. It may also illuminate other related issues, one being the role of interaction energies in many body systems.

In previous work on such cosmologies the focus has been on models with a regular matter configuration of the 4-polytope type. This kind of configuration has been used to simplify the analysis. The regular 4-polytopes are 4-dimensional analogues of the (3-dimensional) Platonic solids. There are six different regular 4-polytopes with 5, 8, 16, 24, 120 or 600 cells. Previous work on such regular models placed identical sources at the cell centres. In this way the regularity of the model could be preserved. The edges of the cells in such a model have a high degree of symmetry in the sense that the gravitational field is locally rotationally symmetric (LRS) about the edge. The edge is then said to be an LRS curve \cite{Clifton_etal:2013}. This symmetry will allow an analytic analysis of the evolution up to some time away from the initial data hypersurface. 

Our primary goal in this work is to relax the obviously unrealistic condition of regularity to see how this will affect the curvature structure and also the evolution. We wish to construct a model in which the sources can be distributed randomly while still preserving at least one LRS curve. It would not be possible to do this with a completely random distribution. However, the existence of LRS curves depends only on discrete symmetries.  Each reflection symmetry has an associated symmetry surface (the mirror). If three or more such symmetry surfaces have a common intersection curve, then that curve is necessarily LRS \cite{Clifton_etal:2017}. Consider 
the regular 4-polytope with 16 cells as an example and fill one of its cells with a random distribution of sources. At an edge of that cell there are in total four cells that meet. One is the cell we have already filled and then there are three more cells meeting at that edge. We can now fill the other three cells by reflecting the given sources in the adjoining faces going around the edge. The newly filled cells can then be used as seeds to fill the rest of the polytope. Since there is an even number of cells at the edges, this procedure gives an unambiguous distribution of sources for the whole polytope. Each cell is then a mirror image of all its neighbours. However the distribution inside the cells can be arbitrarily irregular.

While the above construction works well it cannot be used for any other regular 4-polytope. The reason is that all the other polytopes have an odd number of cells meeting at the edges.  In particular, it cannot be used for the 8-cell polytope which is the one that has been most extensively investigated in the past. For that reason we will use another model with eight identical cells which is also a tessellation of the 3-sphere but which is not of the regular polytope type. Instead, the cells can be characterized as solid  lenses. All the cells meet at a great circle. If the number of cells is of the form $n=6+2k$ with $k$ a non-negative integer, then the common great circle retains the LRS property when the above construction of the source configuration is used. The choice $k=1$ gives us an 8-cell model which contains the desired LRS curve. 

We will
use the second half of the Greek alphabet $(\mu, \nu, \rho\ldots =0,1,2,3)$
for spacetime coordinate indices and the second half of the Latin
alphabet to indicate spatial coordinate indices $(i, j, k\ldots =1,2,3)$.
The first half of the Latin alphabet and the
first half of the Greek alphabet will be used for spacetime orthonormal frame 
indices $(a, b, c\ldots=0,1,2,3)$ and spatial orthonormal frame
indices $(\alpha, \beta, \gamma\ldots= 1,2,3)$ respectively.


\subsection{The Initial Metric of the DI Model}

Following \cite{Clifton_etal:2012} (see also \cite{Bentivegna})  we consider an initial \emph{instantaneously static} spatial hypersurface with metric
\begin{equation}\label{eq:metric}
   h = \psi^4 \hat h_{ij} \textrm{d} x^i \textrm{d} x^j
\end{equation}
where $\hat h$ is the metric of a 3-sphere and $\psi$ is a conformal factor to be specified below. In hyperspherical coordinates the metric has the form
\begin{equation}\label{hat_h}
   \hat h =\textrm{d}\chi^2 + \sin^2\! \chi(\textrm{d}\theta^2 + \sin^2\theta\,\textrm{d}\phi^2)
\end{equation}
with coordinates in the ranges $0<\chi <\pi$, $0<\theta <\pi$ and $0\leq\phi <2\pi$.
For some purposes it is useful to embed the 3-sphere in a 4-dimensional Euclidean space with Cartesian
coordinates $\left(w,\, x,\, y,\, z\right)$. The relations to the hyperspherical coordinates are
\begin{eqnarray*}
w & = & \cos\chi\\
x & = & \sin\chi\sin\theta\cos\phi\\
y & = & \sin\chi\sin\theta\sin\phi\\
z & = & \sin\chi\cos\theta \ .
\end{eqnarray*}
The point where \( \chi = 0\) is referred to as the north pole of the hypersphere.

We will proceed as in \cite{Clifton_etal:2012} and solve
the Gauss-Codazzi equations:
\begin{equation}\label{eq:Gauss-Codazzi}
 \begin{split}
   R+K^{2}-K_{ij}K^{ij}&=0 \\[3pt]
   (K_{i}^{\,\,\,\, j}-\delta_{i}^{\,\,\,\, j}K)_{|j}&=0
 \end{split}
\end{equation}
where $K_{ij}$ is the extrinsic curvature of the 3-space, $K$ is
its trace and $R$ is the Ricci scalar of the (physical) 3-space. The vertical line in equation \eqref{eq:Gauss-Codazzi} denotes the (spatial) covariant derivative of the metric $h$. Since the hypersurface
is instantaneously static, the extrinsic curvature vanishes, $K_{ij}=0$,
and the Gauss-Codazzi equations are then reduced to the very simple form
\begin{equation}
R=0 \ . \label{eq: Gauss codazzi eqn}
\end{equation}
For the metric \eqref{eq:metric}, this condition reduces to the Helmholtz equation on the 3-sphere
\begin{equation}
 \hat{\nabla}^{2}\psi=\tfrac{1}{8}\hat{R}\psi \label{eq: Helmholtz equation}
\end{equation}
where $\hat{R}$ is the 3-sphere Ricci
scalar and $\hat{\nabla}^2$ is the 3-sphere Laplacian. Since the Helmholtz equation is linear in $\psi$
we can form a superposition of single black hole solutions. A conformal factor corresponding to \emph{N}
Schwarzschild masses on the background 3-sphere
can be defined as
\begin{equation}\label{conformal_factor}
   \psi(\chi,\,\theta,\,\phi) =
                  \sum_{k=1}^{N}\frac{\sqrt{m_k}}{\sqrt{2(1-n\cdot n_{k})}}
\end{equation}
where $n$ and $n_k$ are unit vectors in the embedding space.
The location of each source is provided by the vector $n_{k}$ while 
\begin{equation}
   n=(\cos\chi,\,\sin\chi\sin\theta\cos\phi,\,\sin\chi\sin\theta\sin\phi,
        \,\sin\chi\cos\theta) \ .
\end{equation}
The full hypersurface metric is then obtained by inserting the conformal factor (\ref{conformal_factor}) in (\ref{eq:metric}).

\subsection{Locally Rotationally Symmetric Curves}
Locally Rotationally Symmetric (LRS) spacetimes have been defined and investigated rather extensively by several authors following \cite{Ellis}. A spacetime is LRS if the tangent space of every point has a direction about which there is no preferred perpendicular direction defined by the geometry. Here we consider spacetimes which are not LRS everywhere but contain curves of LRS points. Our DI models are constructed with symmetry surfaces whose intersections are LRS curves. Such curves are useful since they allow a significant simplification of the dynamics. Specifically, the dynamics for points on the LRS curves reduces to a system of ODEs driven by a non-autonomous term which can be evaluted (in principle) by a recursion procedure \cite{Clifton_etal:2017}. 
In practice, it has only been possible to evaluate the recursion a few steps before it becomes unwieldy. Nevertheless, even the undriven ODE system gives results that match current numerical work to an accuracy of about one percent \cite{Bentivegna 2}.
 
Suppose that there are $n$ surfaces which intersect
along a curve and that the geometry admits reflection symmetry about each of the surfaces. Now arrange
an orthonormal frame $(\mathbf{e}_{1},\,\mathbf{e}_{2},\,\mathbf{e}_{3})$
with $\mathbf{e}_{1}$ parallel to the curve. There will then be a discrete set of rotations about the curve such that
the frame vectors are transformed as
\begin{eqnarray}
\boldsymbol{\tilde{e}}_{1} & = & \boldsymbol{e}_{1}\nonumber \\
\boldsymbol{\tilde{e}}_{2} & = & \boldsymbol{e}_{2}\cos\phi_{q}-\boldsymbol{e}_{3}\sin\phi_{q}\label{eq: transformation of frame vectors}\\
\boldsymbol{\tilde{e}}_{3} & = & \boldsymbol{e}_{3}\cos\phi_{q}+\boldsymbol{e}_{2}\sin\phi_{q}\nonumber 
\end{eqnarray}
while leaving invariant all tensors picked out by the geometry
\begin{equation}
   T^{\tilde{a}\tilde{b}\tilde{c}\ldots}{}_{\tilde{d}\tilde{e}\tilde{f}\ldots}   
     =T^{abc\ldots}{}_{def\ldots}.\ \label{eq: symmetry condition}
\end{equation}
The rotation angles are given by $\phi_q=2\pi q/n$ where $n \geq2$ is the number of intersecting surfaces and $q=1,\ldots, n-1$. For $n\geq3$ it can be shown that curves associated with such a discrete symmetry group have the LRS property \cite{Clifton_etal:2013} while for $n=2$ this does not follow (see discussion in \cite{Bentivegna} and \cite{Clifton_etal:2017}). 
Vectors and tensors satisfying \eqref{eq: symmetry condition} have the following properties when evaluated along LRS curves:

\begin{enumerate}
\item For a vector $T^{\alpha}$ evaluated in the orthonormal
frame we have \hbox{$T^{2}=T^{3}=0$} while $T^{1}$ remains arbitrary.
Therefore vectors respecting the discrete symmetry group are always parallel to the LRS curve.
\item A rank-2 tensor with orthonormal components $T^{\alpha\beta}$
must be diagonal along an LRS curve with components satisfying $T^{22}=T^{33}$.
\end{enumerate}

\subsection{\label{sub: uniform distribution of masses}A Piecewise Statistically Uniform Distribution of Sources in a 3-Sphere Universe}
We wish to construct a universe in which the sources are distributed as randomly as possible but still contains at least one LRS curve. We can then use the existence of the LRS curve to get estimates of the dynamics of the model. To satisfy the LRS requirement we need a model which has a symmetry subgroup containing at least three distinct reflection isometries. 
The symmetry surfaces in such a model leads to a lattice of cells which are identical up to reflections. Every cell face is then part of a symmetry surface for a reflection. For simplicity, consider a lattice with identical cells. One could then try to define a piecewise random source configuration as follows. First distribute $N$ sources randomly in one of the cells. Then for each face of that cell reflect the given source distribution to a neighbouring cell using the reflection symmetry corresponding to that face.
Further reflections can then be used to distribute the given source configuration to all the cells in the model. For such a construction to work, the relevant symmetry subgroup of the model must be preserved when all the cells have been filled. Otherwise, the LRS property will not in general be preserved. It is straightforward to see that a necessary condition for this is that the number of cells meeting at an edge must be even. The number in question is the third Schl\"afli number. Of the regular models in \cite{Clifton_etal:2012}, only the 16-cell satisfies the requirement with four cells meeting at the edges. It can be shown explicitly that the above construction does work for the 16-cell. 
However, we prefer to use a different model that contains only eight cells and  for which there is a simple algorithm that can be used repeatedly to populate the cells.

\subsubsection{Tessellating the 3-Sphere}

Similarly to \cite{Clifton_etal:2012}, we consider a 3-sphere representing a momentarily static hypersurface in the universe. It is tessellated into a set of empty and identical cells. However, unlike
in \cite{Clifton_etal:2012}, the tessellation in our case will not
correspond to any of the regular polytopes but
will instead be taken to resemble hyperspherical versions of ``orange pieces''.

Suppose we have an orange and intend to share it by cutting it into a number of wedge pieces. Typically one might make the cuts where the black
curves are in figure \ref{fig: orange pieces} which would result
in several manageable pieces. If one ensures that each piece is of
the same size, the number of pieces will be determined by the ``equatorial
width'' $\Delta\phi$.

\begin{figure}
\begin{centering}
\includegraphics[scale=0.5]{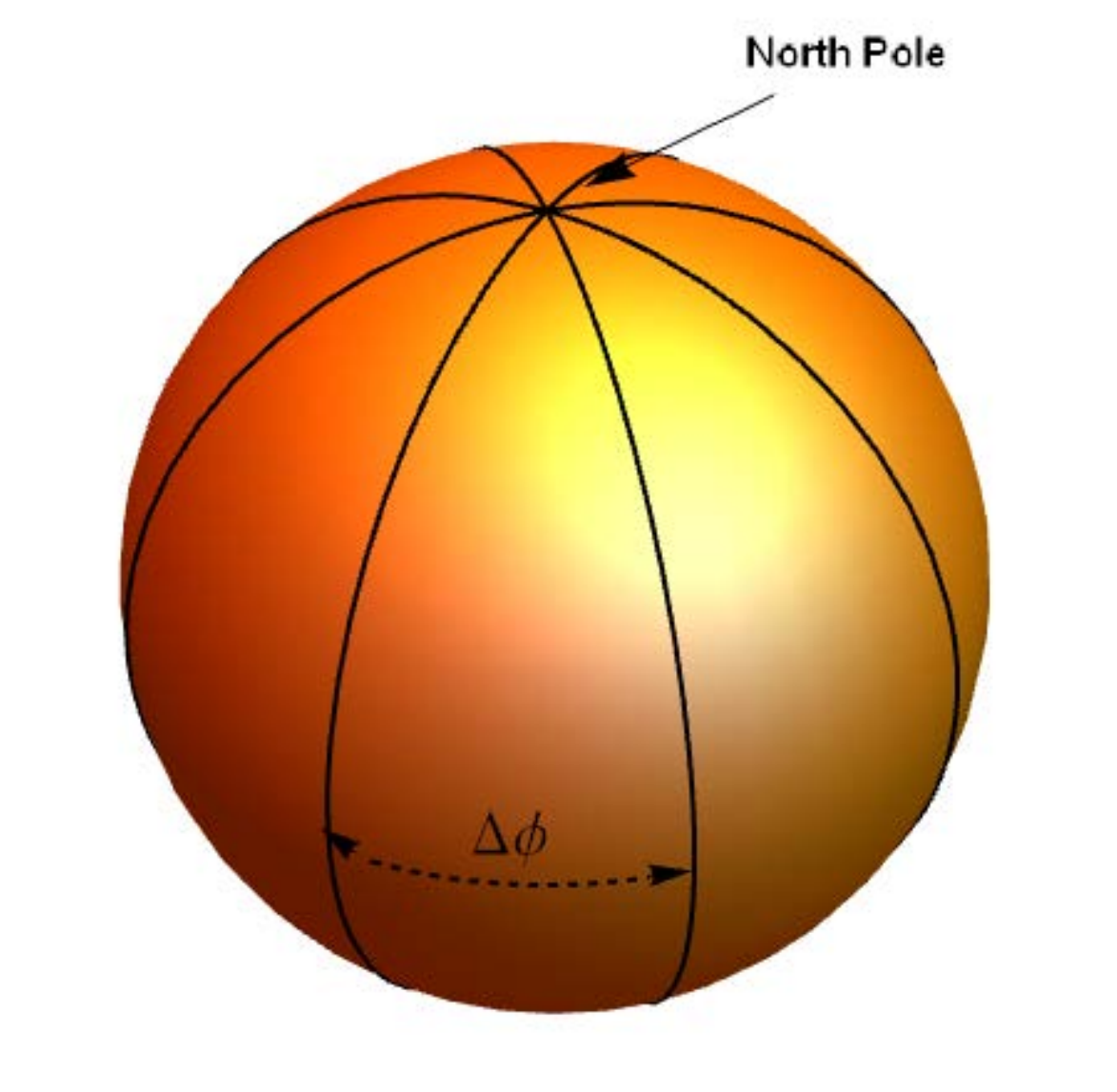}
\par\end{centering}
\caption{\label{fig: orange pieces}If a 2-sphere represents an orange peel, the black curves then represent knife cuts. The surface of each resulting
peel segment is then considered a cell and the cuts are cell boundaries. The dashed line indicates the ``equatorial width'' $\Delta\phi$
which is taken to be the same for all cells. The poles are located opposite
each other at the intersection of the cell boundaries.}
\end{figure}

In terms of a general 2-sphere instead of an orange, the surface of
each orange piece corresponds to a cell, while the knife cuts are
cell boundaries. The cell boundaries are located at fixed $\phi$-coordinates
and the angular distance to the nearest cell boundary is given by
$\Delta\phi$, which must be equal for all cells. In addition, all
cell boundaries intersect at two points - the north and south pole.

Keeping this image in mind, we now make our move to the 3-sphere.
In this transition surfaces become volumes, curves become surfaces
and points become curves. Thus our ``orange'' becomes a 3-sphere,
with 3-dimensional cells, each bounded by two 2-dimensional spherical
surface segments. These cell boundaries intersect along a curve $C_{LRS}$
(instead of at merely two points) which is a great circle passing through the 3-sphere's north and south poles located at
$\chi=0$ and $\chi=\pi$ respectively. In Cartesian coordinates $\left(w,\, x,\, y,\, z\right)$
the curve $C_{LRS}$ is parametrized by
\begin{equation}
   C_{LRS} = \left(\cos\lambda,\,0,\,0,\,\sin\lambda\right) \quad\text{where}
              \quad  0\leq \lambda<2\pi \ .
\end{equation}
The cell boundaries $\partial V_{q}$ are a series of spherical surface segments
which enclose the 3-dimensional cells $V_q$. Using hyperspherical
coordinates $\left(\chi,\,\theta,\,\phi\right)$, the cell boundaries
$\partial V_{q}$ are given by
\begin{equation}\label{eq: location of cell boundaries}
   \partial V_{q}=(\chi,\,\theta,\,\frac{\pi}{p}q) \quad\text{where}\quad
     p\geq3 \quad\text{and}\quad q=0,1,\ldots , 2p-1 \ .
\end{equation}
The integer $p$ corresponds to the number of intersecting surfaces
(cell boundaries). The form of each cell can be viewed as similar to a biconvex spherical lens (cf.\ \cite{Gausmann_etal_2001}).


\subsubsection{\label{sub: Marsaglia's algorithm}Filling the cells}

After tessellating the 3-sphere as described above, the resulting
cells are to be filled with a number of Schwarzschild black holes in a manner which ensures that the
curve $C_{LRS}$ actually remains LRS. 
We do this by first designating one of the cells as a \emph{starting cell} (this
choice is completely arbitrary since all cells are identical). Then
fill the starting cell with one or more points (corresponding to the location of sources, specifically black holes) in any desirable manner.
We choose to distribute the points randomly in the starting cell by using an algorithm invented by Marsaglia \cite{Marsaglia}. The algorithm is designed to generate random uniform distributions on the 3-sphere.
\begin{enumerate}
\item First select four points $V_{1},\, V_{2},\, V_{3}$ and $V_{4}$ randomly
from the interval $[-1,\,1]$ and ensure that $S_{1}=V_{1}^{2}+V_{2}^{2}<1$
as well as $S_{2}=V_{3}^{2}+V_{4}^{2}<1$. 
\item Then the following point is picked randomly from a uniform distribution
on the unit 3-sphere:
\begin{equation}
w=V_{4}\sqrt{\frac{1-S_{1}}{S_{2}}},\qquad x=V_{1},\qquad y=V_{2},\qquad z=V_{3}\sqrt{\frac{1-S_{1}}{S_{2}}}.\label{eq: uniform point on 3-sphere}
\end{equation}
\item If it lies outside the starting cell - discard it. Otherwise - keep
the point. Initially we only want points inside our starting cell.
\item Repeat steps 1-3 until a desirable number of points lie inside the
starting cell.
\end{enumerate}

\begin{figure}
\begin{centering}
\includegraphics[scale=0.45]{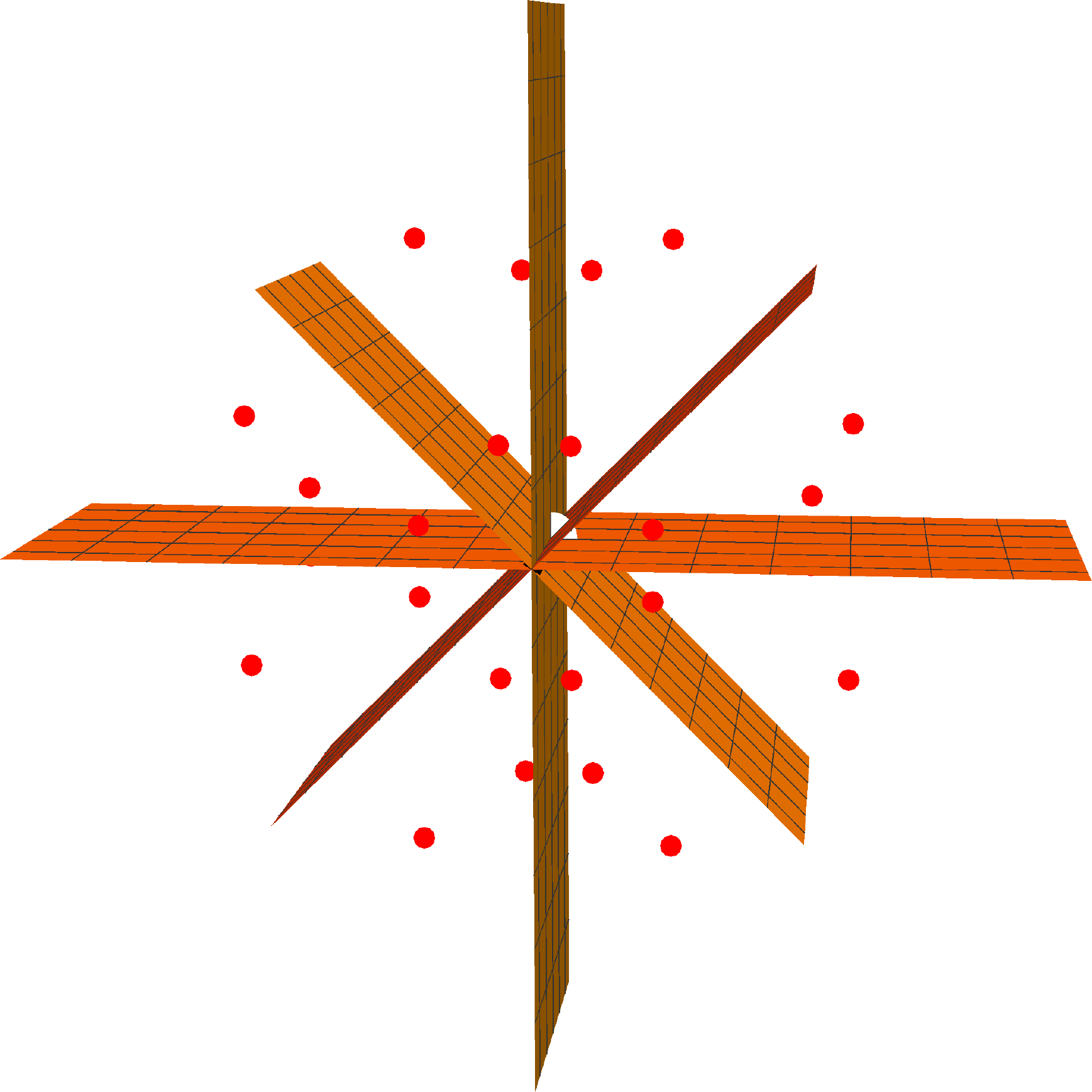}
\par\end{centering}

\caption{\label{fig: Stereographic projection}Stereographic projection of
our 3-sphere tessellated into 8 cells. The planes are the cell
boundaries and the points correspond to black holes while the white
point in the middle indicates the location of the south pole at $\chi=\pi$.
Notice that the contents of each cell is a mirror image of its neighbour,
since the cell boundaries permit reflection symmetry by construction. The planes intersect
each other along a line which corresponds to the LRS curve. Notice
also the discrete rotational symmetries that exist for rotations about the intersection line. To be specific, there are 4 distinct rotations (multiples of $\pi/2$, \emph{cf}.\ \eqref{eq: location of cell boundaries}) about that line leaving the space invariant.}
\end{figure}

After the starting cell has been filled with a random configuration
of points, the remaining cells are filled with a point configuration such that every cell becomes the mirror image of its neighbour.
This is most easily done in the stereographic projection of the 3-sphere,
where the cell boundaries are flat surfaces instead of spherical -
see figure \ref{fig: Stereographic projection}. Since
the cell boundaries are symmetry surfaces, the configuration
of the starting cell can easily be mirrored into the neighbouring cells.
Once every cell is a mirror image of its neighbour, an inverse of the
stereographic projection can be made to return to the original 3-sphere.
The cells will still remain mirror images of each other since the
projection is conformal. Furthermore since two reflections correspond
to a rotation there will also be a discrete rotational symmetry around
the common cell boundary $C_{LRS}$ \cite{Shan}. 

\subsubsection{The Specifications of our Model}
We have chosen to investigate the dynamics of a DI orange piece model consisting of 8 cells corresponding to $p=4$. This is a model with few cells which is also comparatively simple to handle technically when defining the source distributions. In addition, having 8 cells allow us to compare with regular models in \cite{Clifton_etal:2013} in which the total number of masses investigated were 8, 16, 24, 120 or
600 (corresponding to 1, 2, 3, 15 or 75 masses per cell respectively).


\section{Finding the Proper Mass of Schwarzschild Black Holes in the DI Model\label{sec: Proper mass}}

In paper \cite{Clifton_etal:2012}, two types of mass were discussed,
\emph{viz.}\ the effective mass and the proper mass. The effective
mass $m_{\text{eff}}$ corresponds to the mass parameters $m_k$ in \eqref{conformal_factor}. It is the mass the source would have if it were alone in an asymptotically flat universe. 
However when multiple masses are present,  $m_{\text{eff}}$ will also contain interaction energies (which are negative). 
In this paper, the effective mass is set to a constant. 
The reasoning behind this is simplicity - it is easier to calculate proper mass from a fixed effective mass rather than the other way around. 
The actual value of $m_\text{eff}$ will then simply correspond to an overall scaling of all the masses. 
We therefore present the data as the ratio $m_{p}/m_{\text{eff}}$.

The proper mass is defined as in \cite{Clifton_etal:2012} and
is analogous to the concept of bare mass in \cite{Brill and Lindquist}. Physically it can interpreted as the ``local'' mass as measured in a vicinity of the black hole. The difference between the two mass concepts can be interpreted as due to the interaction energies between all the sources in the configuration.
The proper mass of a Schwarzschild black hole at $\chi=0$ (north pole) is calculated
by expanding the conformal factor \eqref{conformal_factor} according to
\begin{equation}
\sum_{k=1}^{N}\frac{\sqrt{m_k}}{\sqrt{2(1-n\cdot n_{k})}}=\frac{2c_{1}}{\chi}+c_{2}+O(\chi)\label{conformal_expansion}
\end{equation}
where $c_{1}$ and $c_{2}$ are constants depending on the mass configuration. The proper mass is then given by \cite{Helena}
\begin{equation}
m_{p}=4c_{1}c_{2}. 
\end{equation}
The proper mass of sources at other positions can be calculated by first rotating the coordinate system (or the mass configuration) to place the north pole at a given source.

Referring to the expansion (\ref{conformal_expansion}) it is clear that the proper mass depends on the configuration of sources -- different configurations will yield different proper mass values. 
Since any source configuration is randomly generated, a wide range of proper masses can be expected. We proceed by randomly generating many (over 200) configurations and calculating the proper mass of the sources. 

However, not all sources in a given configuration need to be accounted for. Any source in one cell and its counterpart in every other cell are equally ``massive'', because each cell are identical up to a rotation or reflection.  It is therefore not necessary to study \emph{all} proper masses in a given configuration, only those inside a single cell.

\begin{figure}
\begin{centering}
\includegraphics[scale=0.6]{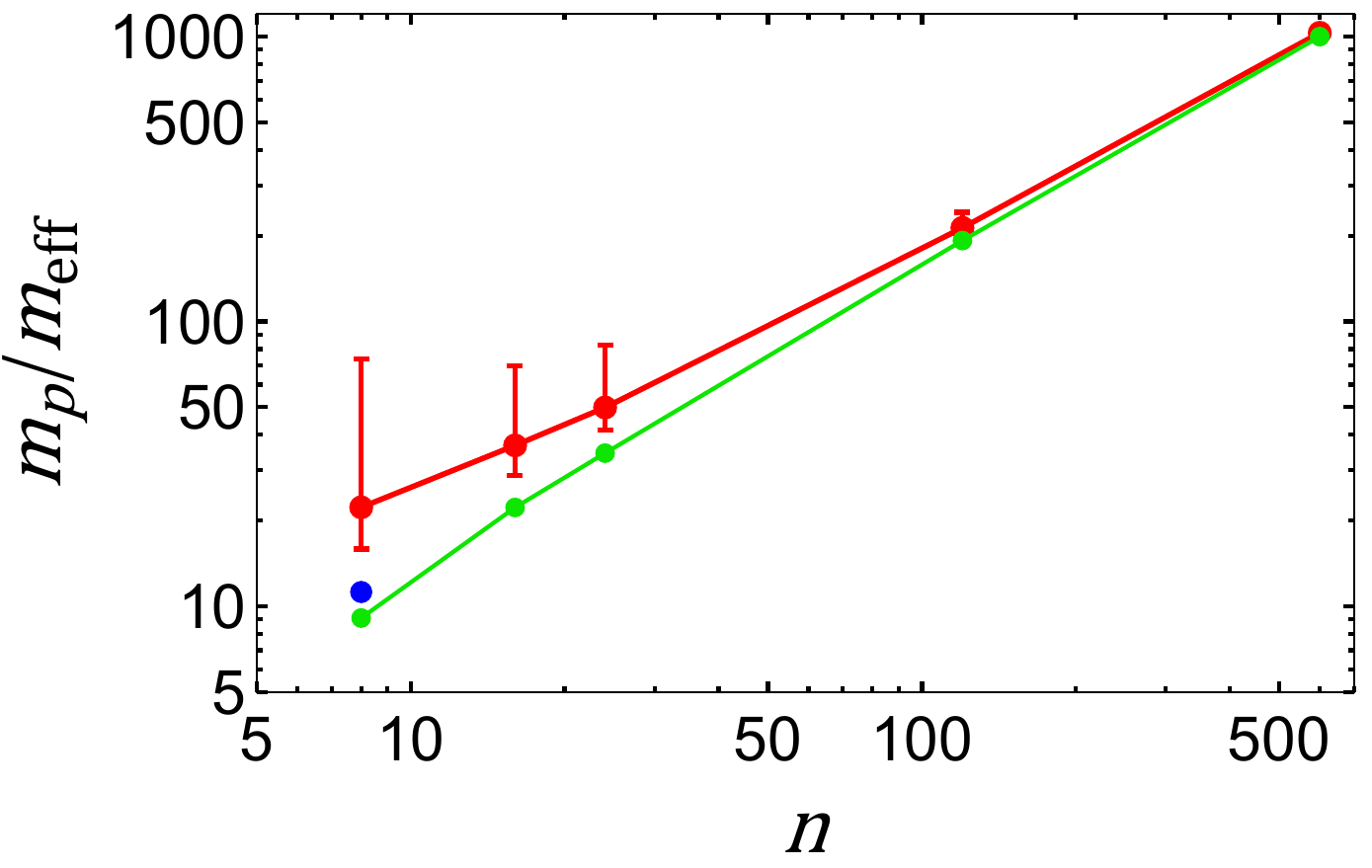}
\par\end{centering}

\caption{\label{fig: Mass ratios}Mass ratios $m_{p}/m_{\text{eff}}$ plotted against the number of masses \emph{n} obtained from more than 200 randomly generated configurations. 
The red curve shows the median mass ratios and the error bars indicate the quintiles. 
For 8 sources, there is the possibility to have one source at the center of each cell. This is the closest our DI model can approximate the regular 8-source model in \cite{Clifton_etal:2012} and is indicated by the blue dot. The green curve depicts the regular ratios found in the aforementioned paper.}
\end{figure}

We have chosen to characterize the distributions by their median, and upper and lower quintiles  (which separates the highest, respectively the lowest, 20\% of the data from the remaining 80\%). These are then plotted as a function of the number of sources $n$, as done in figure \ref{fig: Mass ratios}.
We can also compare with the  mass ratios obtained for the regular models in \cite{Clifton_etal:2012}.
Clearly the mass ratios for the DI and the regular model converge for large $n$ \cite{Clifton_etal:2012,Clifton_etal:2013} and become more massive.
It therefore appears that the overall interaction energies of the DI model are approaching those of the regular configuration in the limit of large $n$.


\section{Initial Behaviour on the LRS Curve}

\subsection{\label{sub: initial length}Initial Length of the LRS Curve}

The initial length of the LRS curve is simple to calculate, since
the curve is a great circle which can be parametrized by $\chi$ in
hyperspherical coordinates. Using the metric given by the expressions
(\ref{eq:metric}) and (\ref{conformal_factor}), the LRS curve
length is then given by the integral
\begin{equation}
L_{0}^{\,\,\, DI}=2 \intop_{0}^{\pi}\psi^{2}(\chi,\,0,\,\phi)\,\mathrm{d}\chi.\label{eq: LRS curve initial length}
\end{equation}
A factor of two arises since both halves of the LRS curve are identical (due to reflection symmetry) and the above integral only takes half of the curve into account.

This can be compared to the
spherical matter-dominated FLRW model (with no dark energy present) at the moment of maximum expansion:
\begin{equation}
\d l^{2}=\frac{16M^{2}}{9\pi^{2}}(\d\chi^{2}+\sin^{2}\mbox{\ensuremath{\theta}}\,\d\theta^{2}+\sin^{2}\chi\sin^{2}\theta\,\d\phi^{2}).
\end{equation}
In this geometry, the length of a great circle with the same parametrizations
as mentioned above is 
\begin{equation}
L_{0}^{\,\,\,\text{FLRW}}=8M/3,\label{eq: length in FLRW}
\end{equation}
where $M$ is the total (proper) mass in the universe.

\begin{figure}[t!]
\begin{centering}
\includegraphics[scale=0.6]{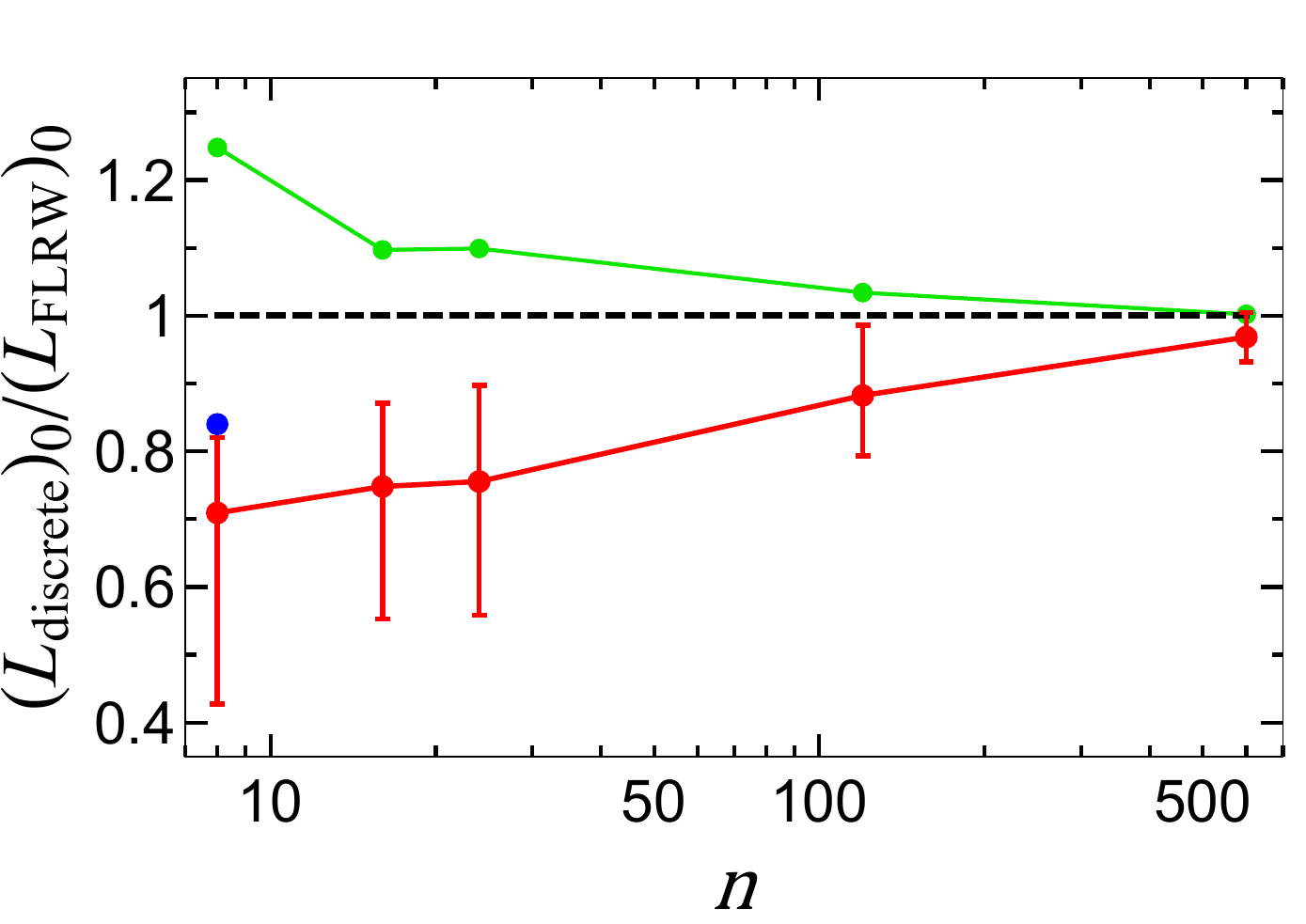}
\par\end{centering}

\caption{\label{fig: LRS curve length plot}Comparison of the (LRS) curve lengths in the DI model and the FLRW model as a function of source count $n$. The red curve depicts the lengths in the
DI model  $(L_{\text{DI}})_{0}$  and error bars indicate the quintiles. The green curve depicts the
length in the regular models $(L_{\text{regular}})_{0}$  in \cite{Clifton_etal:2012}. The dashed horizontal line corresponds
to the FLRW model. The blue dot corresponds to the DI model with a single source at the center of each cell. On the vertical axis - the symbol $(L_{\text{discrete}})_{0}$ is
a placeholder for either $(L_{\text{regular}})_{0}$ or $(L_{\text{DI}})_{0}$. The zero subscript indicates that the lengths are calculated on the initially time-symmetric hypersurface at $t=0$.}
\end{figure}

Figure \ref{fig: LRS curve length plot} plots the ratio between the LRS curve length in the DI model and the FLRW model as a function of the number of sources.
The length in the FLRW model is computed according to equation \eqref{eq: length in FLRW} where $M$ is the total proper mass in the DI model (obtained using the method in section \ref{sec: Proper mass}).
The large dots indicate the median, while the accompanying error bars indicate quintiles.
For histograms detailing the length distributions in detail, see \cite{Shan}.
By including data for the regular models from \cite{Clifton_etal:2012}, we see that both model types approach the FLRW limit as more sources are added.
Our results are so far consistent with \cite{Korzynski_backreaction_continuum_limit}, where a thorough analysis is made of the continuum limit.
However, while the regular models approach the FLRW limit from above, the irregular models approach it from below. 
There are thus two different ways to approach the limiting homogenous case presented by the FLRW model. 

A possible cause for these two types of behaviours can be found by examining figure 8 in \cite{Durk_Clifton}. 
In that work, a cosmological model with clusters of black holes and a tessellation identical to that in \cite{Clifton_etal:2012} is studied. 
There the authors also observe an approach to the FLRW limit from above.
This observation suggests that the choice of tessellation of the 3-sphere will affect how the FLRW limit is approached.


\subsection{Initial Curvature Along the LRS Curve \label{sub: Initial-Curvature-Along}}
The DI models, being vacuum universes only support Weyl curvature. This is in contrast to FLRW models which have only Ricci curvature as a consequence of their conformally flat nature. The Weyl curvature tensor $C_{\mu\nu\rho\sigma}$ can be invariantly decomposed in two parts, a gravito-electric part $E_{\mu\nu}$ and a gravito-magnetic part $H_{\mu\nu}$. They are defined by the relations
\begin{equation}
   E_{\mu\nu}:=C_{\mu\rho\nu\sigma}u^{\rho}u^{\sigma} \ ,\qquad
   H_{\mu\nu}:=\ce{\space}^{*\nhalf}C_{\mu\rho\nu\sigma}u^{\rho}u^{\sigma}
\end{equation}
where $u^{\mu}$ is a 4-velocity field representing an observer family and ${}^{*\nhalf}C_{\mu\rho\nu\sigma} u^{\rho} u^{\sigma}$ is the dual Weyl tensor. The latter is given by
\begin{equation}
   {}^{*\nhalf}C_{\mu\nu\rho\sigma}
    := \tfrac12 \eta_{\mu\nu}{}^{\kappa\tau} C_{\kappa\tau\rho\sigma}
\end{equation}
where $\eta_{\mu\nu\rho\sigma}$ is the Levi-Civita tensor. The tensors $E_{\mu\nu}$ and $H_{\mu\nu}$ are symmetric and traceless by their definition. Their definition also implies the relations $E_{\mu\nu}u^\nu =0$ and $H_{\mu\nu} u^\nu =0$. This shows that they are spatial tensors lying in the rest spaces of the $u^\mu$ observers.
At the initial momentarily static hypersurface, the gravito-magnetic field is zero while the gravito-electric field equals the traceless part of the intrinsic Ricci 3-curvature of the initial hypersurface, $E_{\mu\nu} = {}^{(3)\!}R_{\mu\nu}$ (see e.g.\ \cite{Clifton_etal:2013}).

We follow the convention \cite{van_Elst&Uggla:1997} and define the following five variables representing the orthonormal components of any symmetric tracefree spatial 
\hbox{rank-2} tensor (using the gravito-electric tensor $E_{\alpha\beta}$
as an example): $E_{+}=-\frac{3}{2}E_{11}$, $E_{-}=\frac{\sqrt{3}}{2}(E_{22}-E_{33})$,
$E_{1}=\sqrt{3}E_{23}$, $E_{2}=\sqrt{3}E_{31}$ and $E_{3}=\sqrt{3}E_{12}$.
However, taking ${\bf e}_1$ to be directed along the LRS curve, tracefree symmetric rank-2 tensors are diagonal with $E_-=0$. Therefore the
quantity $E_{+}$ is the only nonzero curvature component along the LRS curve.
We will investigate the behaviour of 
$(E_{+})_0$ along the LRS curve on the initial static hypersurface with the zero  subscript indicating evaluation at $t=0$.

The LRS curve, as we have defined it, is located at $\theta=0,\,\pi$. However, as is evident from Eq.\ 1, the determinant of the metric is zero where $\theta=0,\,\pi$ implying that the entire curve is located at a coordinate
singularity. Therefore, to calculate the curvature, we first rotate the
configuration such that the LRS curve instead lies at $\chi=\theta=\pi/2$
and is parametrized by $\phi\in[0,\,2\pi]$. Therefore in the ensuing
graphs, the position on the LRS curve is given by the hyperspherical
coordinate $\phi$.

Figures \ref{fig: E+ for 8, 16 and 24 masses} - \ref{fig: E+ for 600 masses}
show some examples of how the quantity $(E_{+})_{0}$ behaves along
the LRS curve. It can be seen that in some cases the sign of $(E_{+})_{0}$
changes. To compare with the regular models of Ref.\ \cite{Clifton_etal:2013} we can consider the sign of $(E_+)_0$ along the edges of the cells in that paper, noting that it is negative. Further remarks about the sign distribution can be found at the end of this section. 
We will also return to discuss a physical interpretation of the sign of $(E_+)_0$ when treating time evolution in Sect.~\ref{sect:time}

Now let us take a look at each case individually. From figure \ref{fig: E+ for 8, 16 and 24 masses} we see a unique behaviour exhibited when only 8 sources are present, namely the curvature is constant! This is always the case for any configuration
consisting of only 8 sources: the curvature is constant with $(E_{+})_{0}<0$.
The reason behind the constant curvature
is most likely due to the topology of the 3-sphere.

\begin{figure}
\begin{centering}
\includegraphics[scale=0.6]{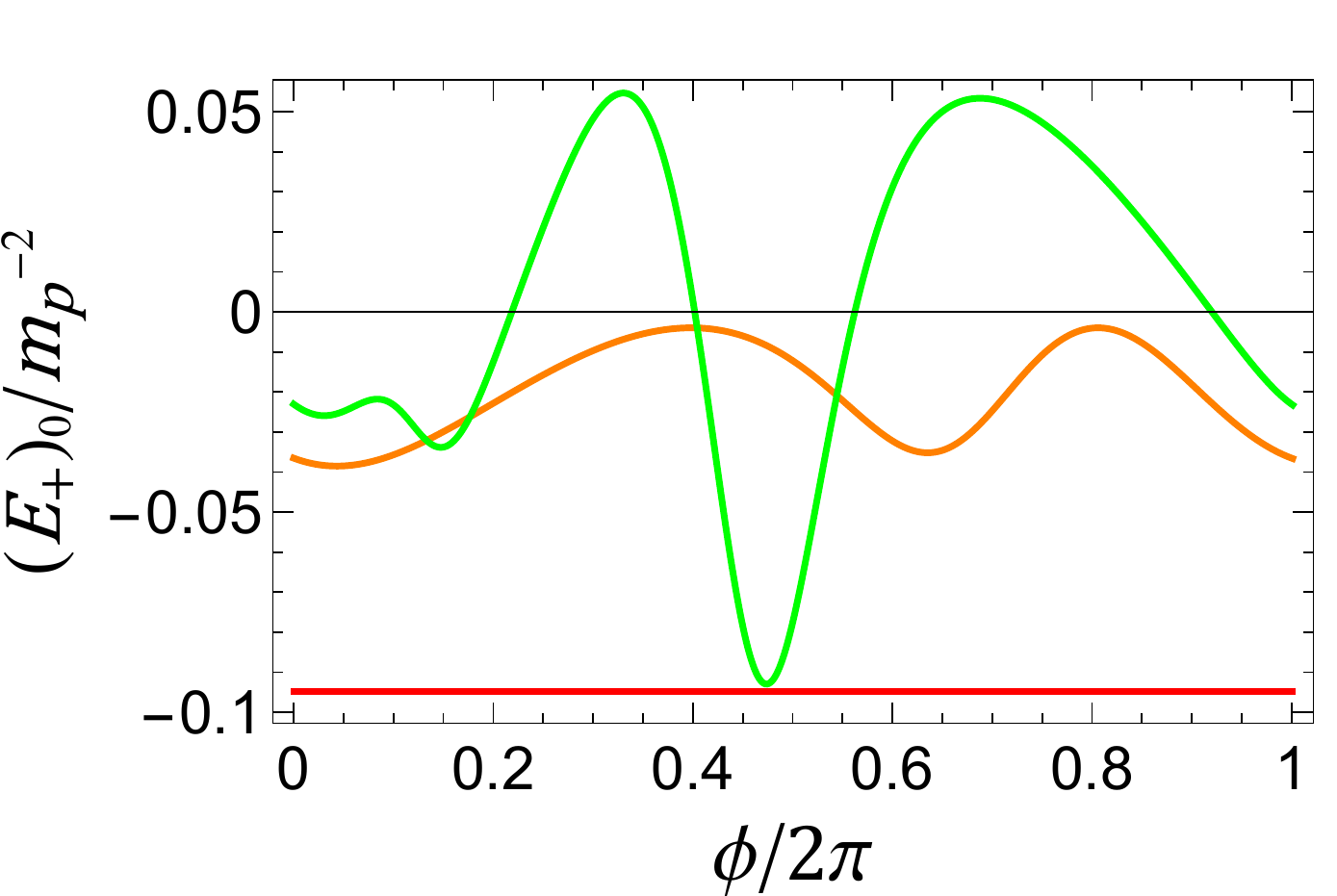}
\par\end{centering}

\caption{\label{fig: E+ for 8, 16 and 24 masses} The curvature $(E_{+})_{0}$
along the LRS curve for three example configurations. Each curve corresponds to a single configuration with 8 (red), 16 (orange) and 24 (green) sources respectively. The curvature is scaled by the inverse square of the median proper mass $m_{p}$ of the corresponding configuration. }
\end{figure}

For 16 and 24 masses, the behaviour of $(E_{+})_{0}$ becomes more
mixed - maxima and minima with occasional changes in the sign
of $(E_{+})_{0}$ are present. Furthermore the behaviour varies with
different configurations. However, something which is consistent among different
configurations is the number of local maxima and minima. For
16 sources the number of maxima found on the LRS curve is two, which is the number of masses per cell. Similarly for
24 sources the number of maxima is three - also the number of masses per
cell. This is evident in figure \ref{fig: E+ for 8, 16 and 24 masses}.
Due to the relatively small number of sources, the effect of each black hole is distinctly visible.

\begin{figure}
\begin{centering}
\includegraphics[scale=0.7]{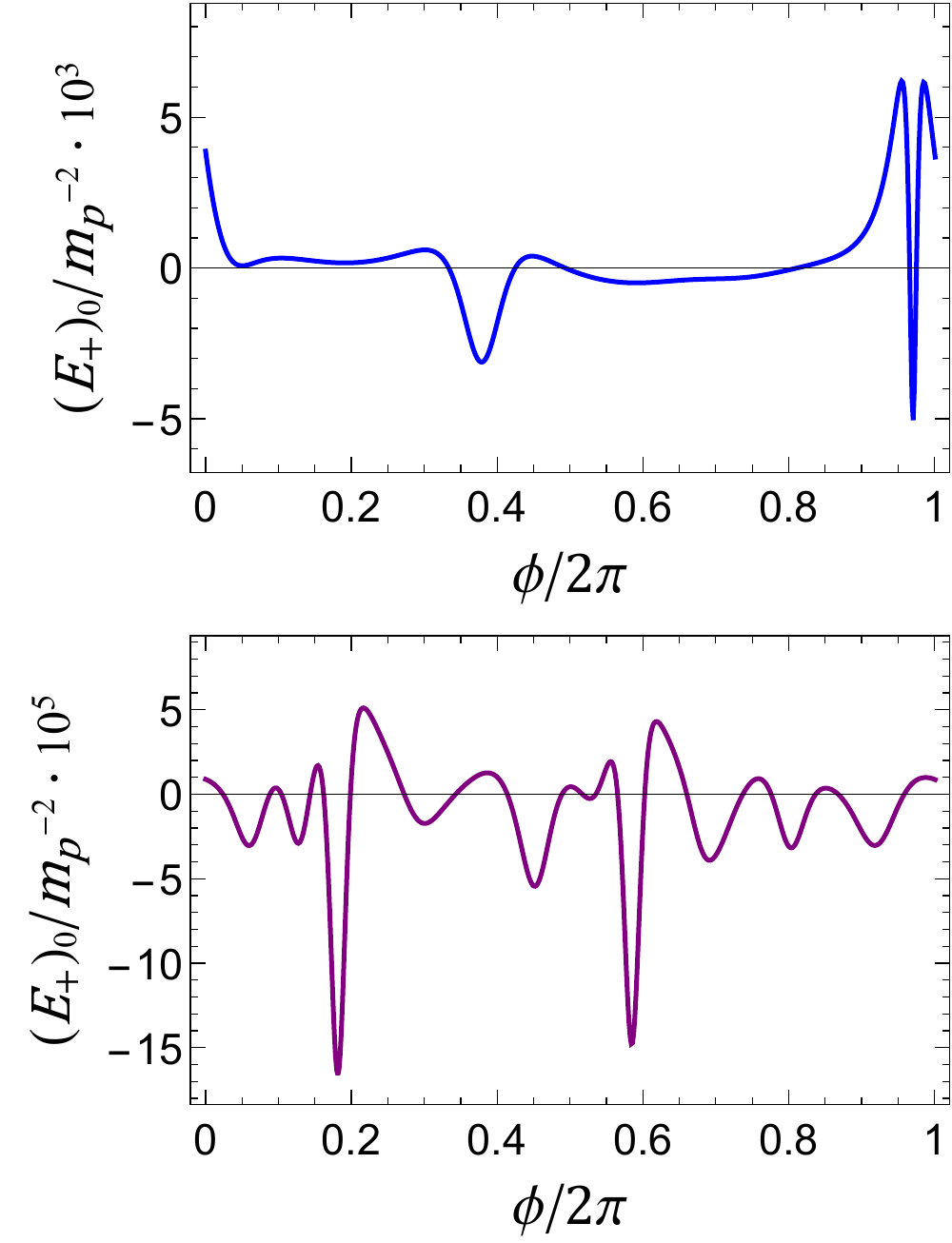}
\par\end{centering}

\caption{\label{fig: E+ for 600 masses} The curvature $(E_{+})_{0}$
along the LRS curve for an example case with 120 sources (top) and 600 sources (bottom). The values are scaled by the median proper mass $m_{p}$.}
\end{figure}

For 120 and 600 masses, the $(E_{+})_{0}$ behaviour is even more
dramatic and irregular. Figure \ref{fig: E+ for 600 masses} 
displays positive peaks and negative troughs and Riemann flat points
at $(E_{+})_{0}=0$. The latter points are characterized by a vanishing
Riemann tensor. The spike-like features are most likely due to a nearby black hole horizon, cf. figure 13 in \cite{Clifton_etal:2013}.

\begin{figure}
\begin{centering}
\includegraphics[scale=0.6]{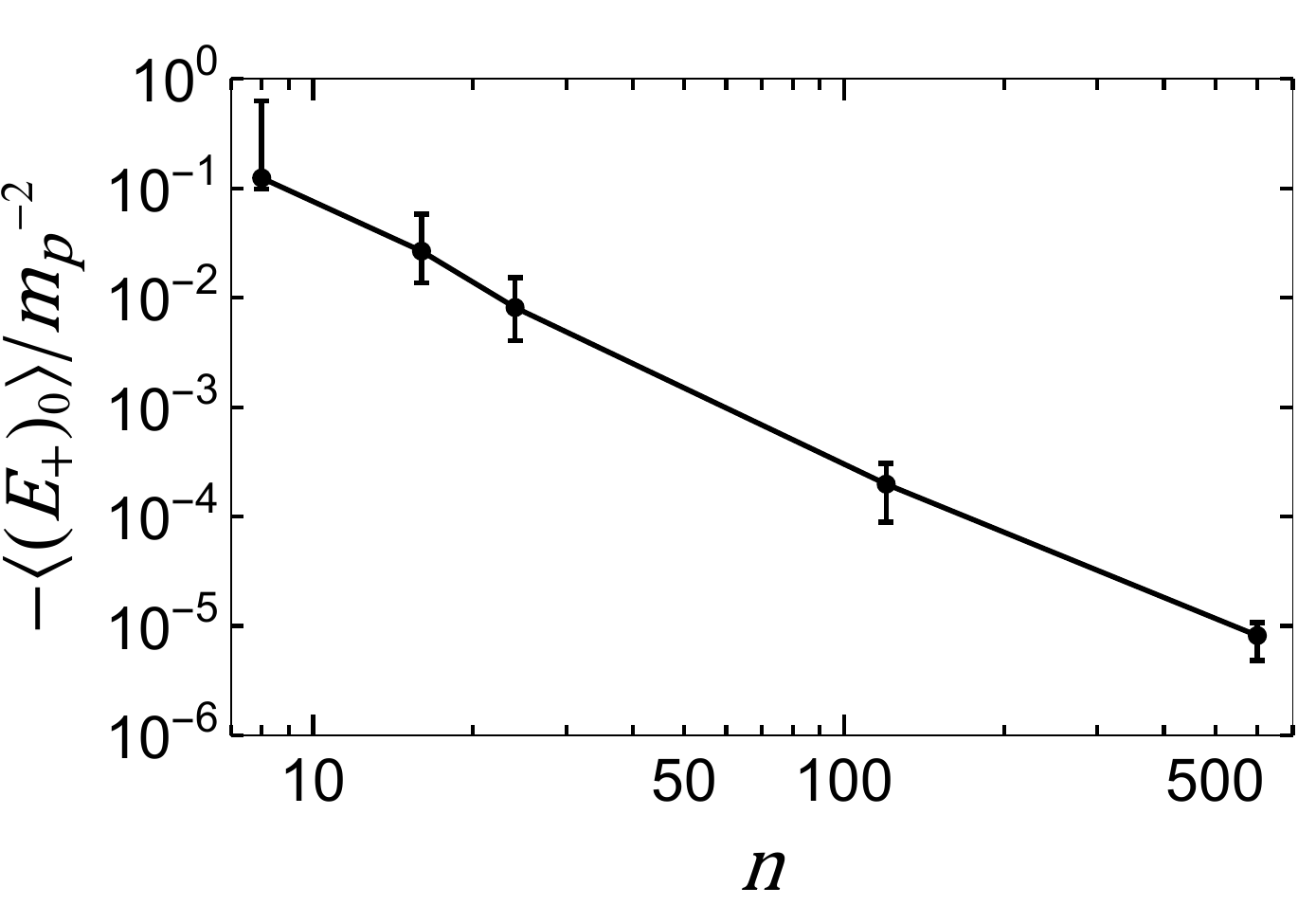}
\par\end{centering}

\caption{\label{fig: median E+}The curvature averaged along the LRS curve 
$\langle(E_{+})_{0}\rangle$ in units of inverse median proper mass squared. This average is  not unique for a given number of sources $n$, but depends on the configuration. The plot above presents the result after generating many random configurations, which yields a distribution. As usual in this paper, dots and error bars  indicate \textit{median} and quintiles of this distribution, respectively.}
\end{figure}

Finally we investigate the sign dominance of $(E_{+})_{0}$ on the
LRS curve. The sign of $(E_{+})_{0}$ at a given point affects the time evolution of any point along the curve. Since there are different signs along the same curve,
it follows that different regions might evolve differently \cite{Clifton_etal:2013}.
A very rough estimate of the sign dominance is provided by the sign of the mean $\langle(E_{+})_{0}\rangle$. Figure \ref{fig: median E+} presents the median
and the location of quintiles obtained from $\langle(E_{+})_{0}\rangle$, as a function of the number of masses. 
Negative values are clearly the most prevalent, approaching $\langle(E_{+})_{0}\rangle$=0 when there are many sources in the model.


\section{Time Evolution}\label{sect:time}

\subsection{Equations Governing Dynamics \label{sub: Equations-Governing-Dynamics}}

With the initial conditions studied, we now proceed to evolve them. We restrict attention to the LRS curve for which the evolution can be studied by analytical methods. The evolution is governed by the system of ordinary differential equations \cite{Clifton_etal:2013,Clifton_etal:2017}
\begin{align}\label{dot_apar}
         \frac{\ddot{a}_\parallel}{\apar} &= \tfrac{2}{3}E_{+} \\[3pt]
   \label{dot_aperp}
   \frac{\ddot{a}_\perp}{\aperp} &= -\tfrac{1}{3}E_{+} \\[3pt]
   \label{dot_E}
   \dot{E}_{+}+3\frac{\dot{a}_{\perp}}{\aperp}E_{+} &= C_+(t)
\end{align}
where
\begin{equation}\label{Cplus}
    C_+(t) = -\tfrac32 n^a n^b \curlH_{ab} \ .
\end{equation}
Here $n^a$ is a unit vector along the LRS curve and 
\begin{equation}
   \curlH_{ab} = \epsilon_{cd(a} D^c H^d{}_{b)}
\end{equation}
and $D_a$ is the spatial covariant derivative. The quantities $\apar$ and $a_{\perp}$ are the scale
factors parallel and perpendicular to the LRS curve. Although the curl\emph{H} term depends on spatial derivatives it can be computed recursively term by term as a Taylor series in $t$ (for given initial time-symmetric data) using the full Einstein vacuum equations (see \cite{Clifton_etal:2017} for details). Therefore, the system \eqref{dot_apar}--\eqref{dot_E} can be considered as an autonomous ODE system driven by the explicitly time-dependent curl\emph{H} term.
The initial values of the scale factors on the time-symmetric hypersurface (at $t=0$) can be taken to be $(\apar)_0 =(\aperp)_0 =1$ and the time-symmetry implies $\dot{a}_\parallel =\dot{a}_{\perp}=0$. The initial values of the gravitoelectric field is given for each spatial position by $(E_+)_0= -\frac32 \bigl[{}^{(3)}\!R_{11}\bigr]_0$.

Considering first the vertex points in the regular models (see \cite{Clifton_etal:2013}) we note that they are points of local spherical symmetry. Then any tensor object picked out by the geometry can only have a scalar part at the vertex. This implies in particular that $E_{ab} = H_{ab} = 0$. Therefore, the vertex points can be characterized as being Riemann flat. It then follows from Eqs.\ \eqref{dot_apar}--\eqref{dot_aperp} that the scale factors are constant for all time, $a_\parallel(t) = a_\perp(t) 
= 1$.

Next we note that $\curlH_{ab}$ is an odd function of $t$ and that the Taylor expansion of \eqref{Cplus} takes the form
\begin{equation}\label{CTaylor}
   C_+(t) = \sum_{n=1}^\infty \frac{C_+^{(2n+1)}}{(2n+1)!} t^{2n+1}
\end{equation}
where the first nonzero term is of order $t^3$ \cite{Clifton_etal:2017}. In \cite{Clifton_etal:2013}, Clifton \emph{et al.}\ analyzed the autonomous part of the system \eqref{dot_apar}--\eqref{dot_E} (i.e.\ with $C_+(t)$ assumed to vanish identically) for all the regular lattice configurations. By performing a numerical integration of the full Einstein system, Korzy\'nski \emph{et al}.\ \cite{Bentivegna 2} showed that the error incurred by neglecting the $\curlH$ term in \eqref{dot_E} is about 1\% at $t \sim 3m_p$ when evolving $a_\parallel$ and $a_\perp$ at an edge midpoint in an 8-cell lattice. An analytic expression for the first term in \eqref{CTaylor} was given in \cite{Bentivegna 2} as a function of the conformal factor with respect to flat space. Covariant expressions for the first two terms were given in \cite{Clifton_etal:2017}, the first one being
\begin{equation}
   C_+^{(3)} = -6\, n^a n^b \curl\curl (E^2)_{ab}|_{t=0}
\end{equation}
where $(E^2)_{ab} = E_{ac} E^c{}_b$. 
To analyse our model, we will exploit the analytical solution of the undriven part of the system \eqref{dot_apar}--\eqref{dot_E} given in \cite{Clifton_etal:2013}. In Sect. \ref{calculating curlH}, we will estimate the error incurred by neglection of the driving term. From the physical point of view, this approximation amounts to neglecting the gravitomagnetic effects, i.e.\ the part of the gravitational interaction which emanates from the relative motion of the masses.

Turning now to the case of nonzero curvature, we note that the undriven form of \eqref{dot_E} is solved by $E_+ = (E_+)_0 /\aperp^3$. When $(E_{+})_{0}>0$, the scale factors can be expressed in terms
of conformal time $\eta$ by
\begin{equation}
 \begin{split}
   a_{\perp} &=\cos^{2}\eta \\[3pt]
      \apar &=\tfrac{1}{2}(3-\cos^{2}\eta+3\eta\tan\eta) \\[3pt]
     t-t_{0} &=\frac{1}{\sqrt{\frac{2}{3}(E_{+})_{0}}}
                (\eta+\tfrac{1}{2}\sin 2\eta)
 \end{split}
\end{equation}
where $-\frac{\pi}{2}<\eta<\frac{\pi}{2}$  and $t_{0}$ is the constant
of integration.
In the case  $(E_{+})_{0}<0$, the solution is given by
\begin{equation}\label{evol_neg}
 \begin{split}
   a_{\perp} &= \cosh^{2}\eta \\[3pt]
      \apar &= \tfrac{1}{2}(3-\cosh^{2}\eta-3\eta\tanh\eta) \\[3pt]
     t-t_{0} &= \frac{1}{\sqrt{-\frac{2}{3}(E_{+})_{0}}}
                (\eta+\tfrac{1}{2}\sinh 2\eta) \ .
 \end{split}
\end{equation}
where $-\frac{\pi}{2}<\eta<\frac{\pi}{2}$.


\subsection{Evolving $E_{+}$ on the LRS Curve}

The equations given in section \ref{sub: Equations-Governing-Dynamics}
allow us to study the curvature, measured by $E_{+}$, as a function of time. 
A first example is given in figure \ref{fig: E+ evolution for 8 masses},
where we evolve a model with 8 sources. In that case, since the curvature is constant along the LRS curve (see figure \ref{fig: E+ for 8, 16 and 24 masses}), it is sufficient to evolve a single point on the curve. 
The value of $E_{+}$ increases monotonically  until a coordinate singularity appears along 
the entire LRS curve simultaneously, beyond which our current formalism provides no information.

\begin{figure}
\begin{centering}
\includegraphics[scale=0.6]{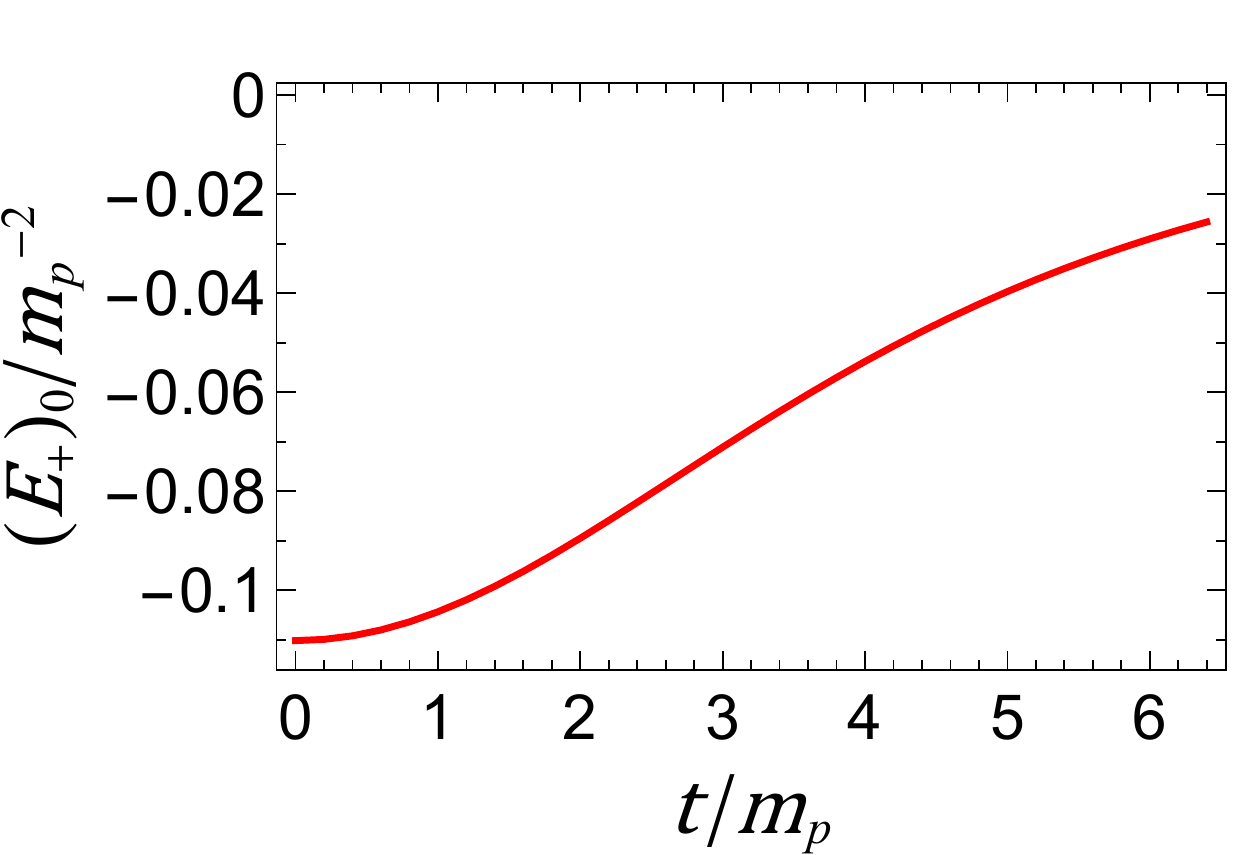}
\par\end{centering}
\caption{\label{fig: E+ evolution for 8 masses}The time-evolution of $E_{+}$ for 8 masses at a fixed point on the LRS curve.
The initial value $(E_{+})_{0}$ is given in figure
\ref{fig: E+ for 8, 16 and 24 masses}. Since the curvature is constant in this case, it is sufficient to study the behaviour at a single point on the LRS curve (see text). Both axes are scaled by the median proper mass $m_{p}$ of the configuration.}
\end{figure}

We have evolved the models with 16 and 24 sources for which initial data was plotted in figure \ref{fig: E+ for 8, 16 and 24 masses}. The result is displayed in figure
\ref{fig: density-plot for 24 masses}. The time and the curvature
$E_{+}$ in those figures are scaled by appropriate powers of the median proper mass $m_{p}$. All points on the LRS curve will eventually
encounter a singularity unless initially Riemann flat (i.e.
$(E_{+})_{0}=0$). The sign of the curvature does not change along the part of the evolution that we can follow.

\begin{figure}
\begin{centering}
\includegraphics[scale=0.7]{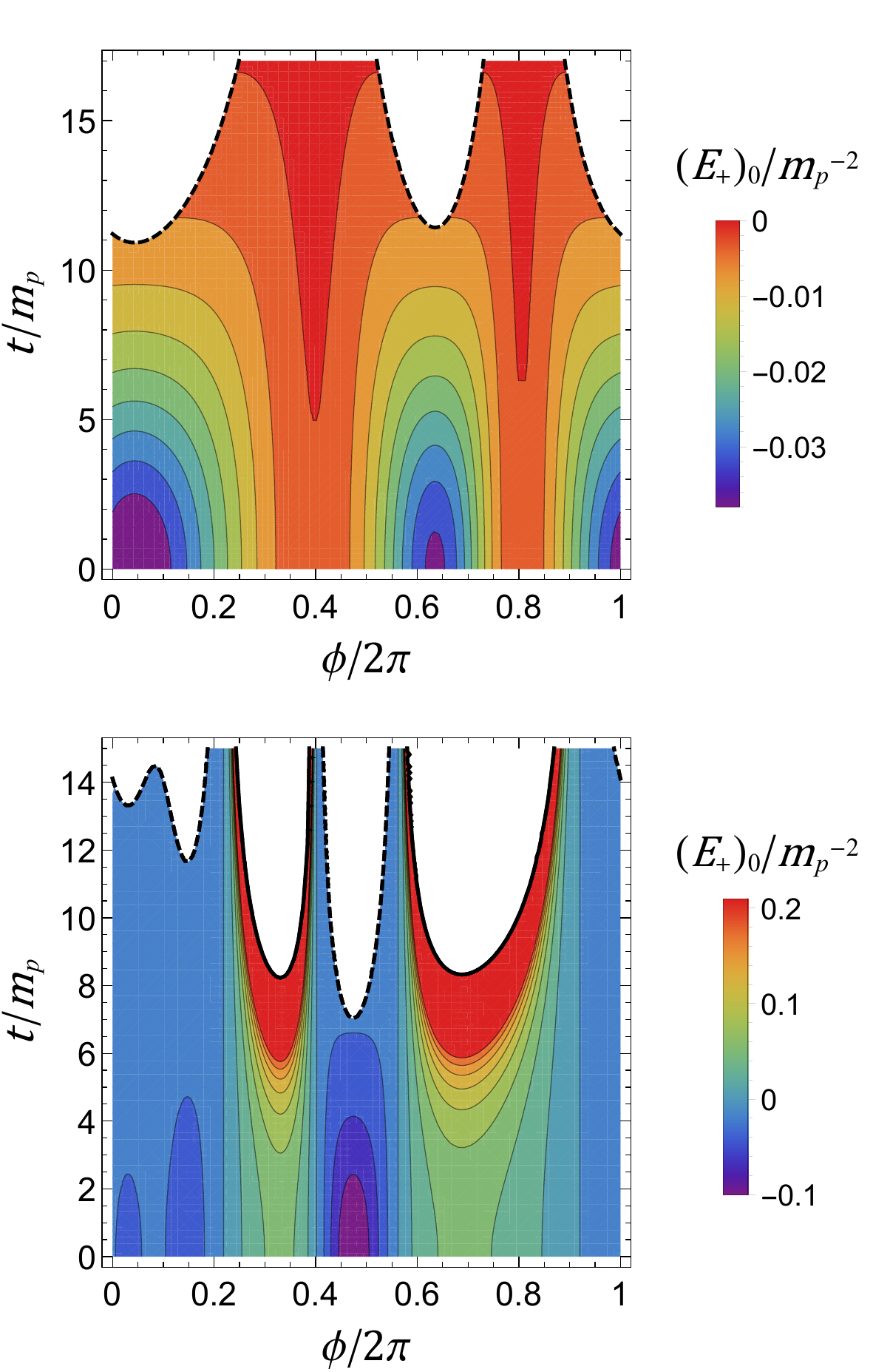}
\par\end{centering}

\caption{\label{fig: density-plot for 24 masses} Density plot of the curvature
for a configuration of 16 sources (top) and 24 sources (bottom). Here, $\phi$ is the angular position
on the LRS curve (a great circle).  Dashed
black curves indicate the location of coordinate singularities, while
solid black curves indicate curvature singularities. See figure \ref{fig: E+ for 8, 16 and 24 masses} for a plot of the initial curvature of this configuration. }
\end{figure}

Referring to the evolution of models with 8, 16 and 24 sources (see figure 
\ref{fig: density-plot for 24 masses}) it appears that coordinate singularities appears before curvature singularities. 
This turns out to be the most common behaviour for the models that we have investigated. See figure \ref{fig: singularity statistics} in which the first appearance of either a coordinate
or curvature singularity is illustrated.

\begin{figure}
\begin{centering}
\includegraphics[scale=0.6]{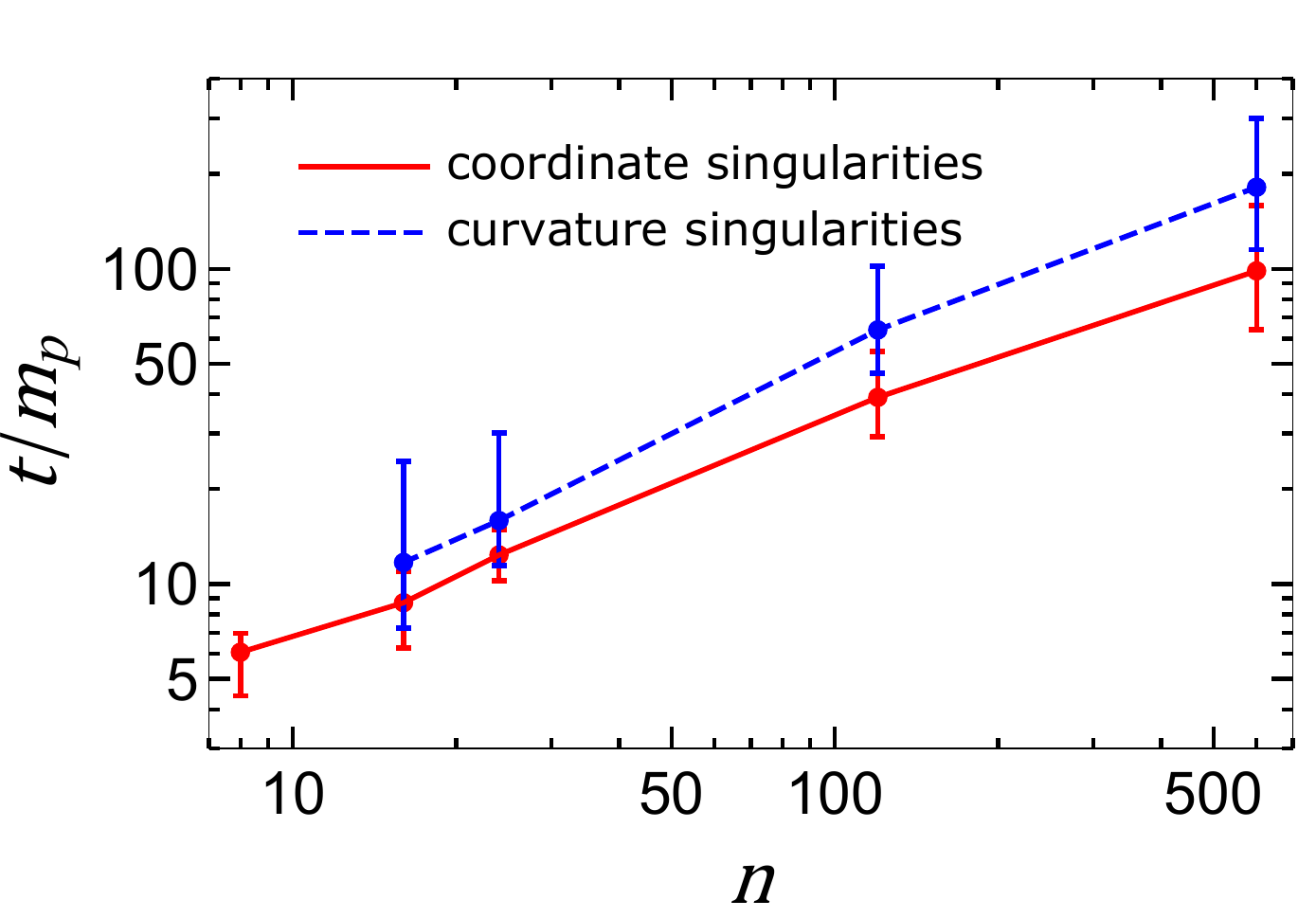}
\par\end{centering}

\caption{\label{fig: singularity statistics}The median time it takes for a
singularity to arise somewhere on the LRS curve as a function of the
number of masses $n$. The error bars indicate as usual the quintiles.}
\end{figure}


\subsection{Dynamic Contractions and Expansions of the LRS Curve}

We now proceed to study the length of the LRS curve 
(i.e. the LRS circumference) as a function of time. It is calculated by
\begin{equation}\label{eq: time dependence of LRS length}
 \begin{split}
   L(t) &= \intop_0^{\pi}\apar(t)\psi^2(\chi,\,0,\,\phi)\,\d\chi
            +\intop_0^{\pi}\apar(t)\psi^2(\chi,\,\pi,\,\phi)\,\d\chi
             \qquad (\theta=0, \pi) \\[5pt]
   L(t) &= \intop_0^{2\pi}\apar(t)\psi^2(\pi/2,\,\pi/2,\,\phi)\,\d\phi
             \qquad (\chi=\theta=\pi/2)
 \end{split}
\end{equation}
which is essentially equation (\ref{eq: LRS curve initial length})
multiplied by a scale factor. Note that $\apar(t)$ has an implicit position dependence as described in section \ref{sub: Equations-Governing-Dynamics}. 
In particular, depending on the sign of $(E_{+})_{0}$, the scale factors will a have qualitatively different time evolution.

The LRS circumference is plotted as a function of time in figure \ref{fig: GCevolution_8} for 500 randomly generated models containing 8 sources.
The lengths are scaled by the circumference $(L_{FLRW})_0$ of a
great circle in a spherical FLRW model with the same total
proper mass $M_{p}$. They are evolved until
a singularity appears.
The similarity in the form of the curves reflects the scaling of time caused by different initial values of $E_+$ in \eqref{evol_neg}. The density distribution of the curves can be attributed to both the dependence on the position of the (single) source in the cells and to the nonlinear (square root) dependence on $(E_+)_0$ in \eqref{evol_neg}.

\begin{figure}
\begin{centering}
\includegraphics[scale=0.6]{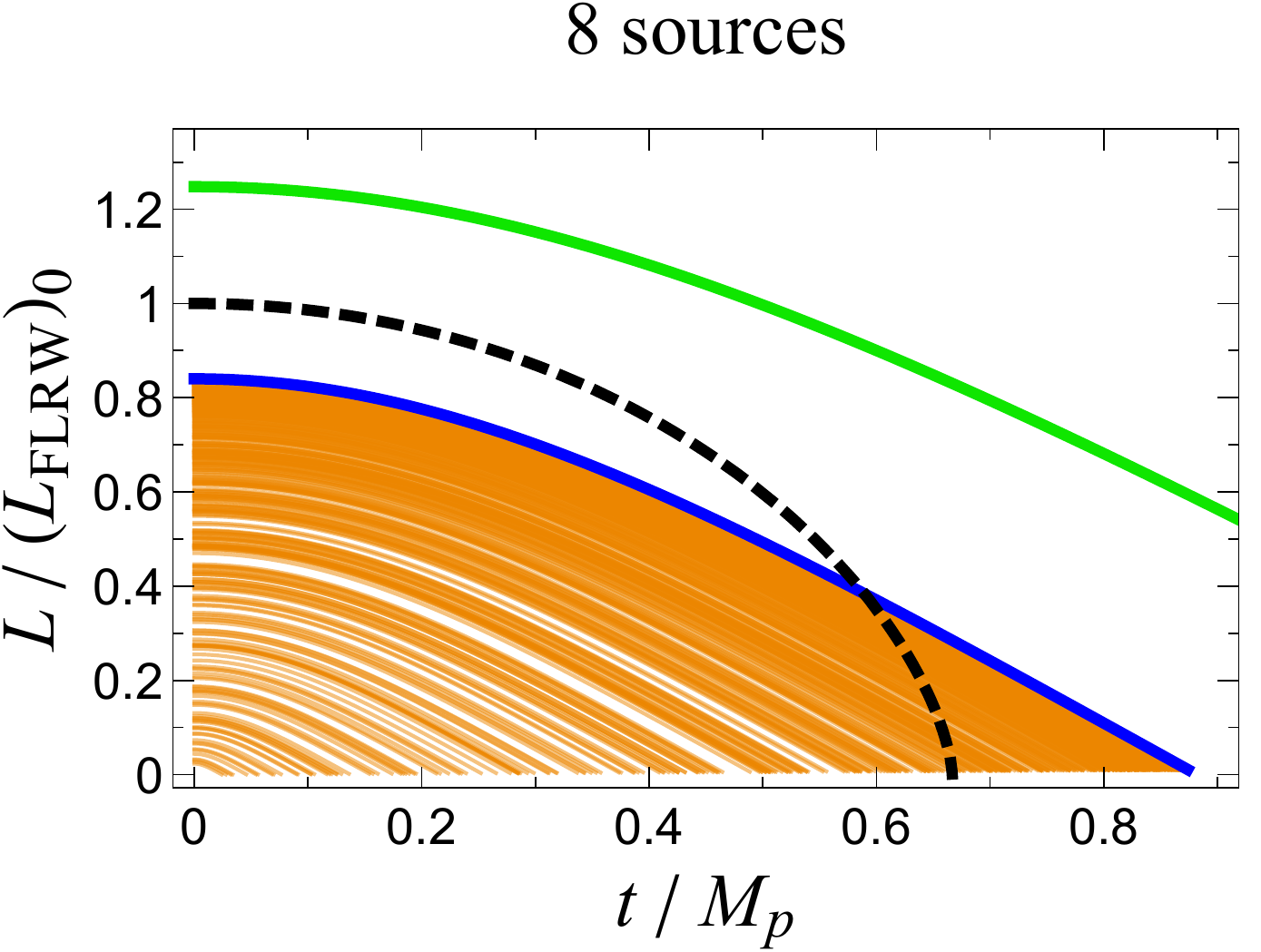}

\smallcaption{\label{fig: GCevolution_8} The circumference of 500 LRS curves as a function of time for 8 sources. Each orange curve corresponds to a randomly generated DI model. The blue curve illustrates the model with sources located at cell centers. The green and dashed black curve depict the regular model with 8 sources given in \cite{Clifton_etal:2013} and the spherical matter dominated FLRW model respectively.}

\par\end{centering}
\end{figure}

\begin{figure*}
\begin{centering}
\includegraphics[scale=0.6]{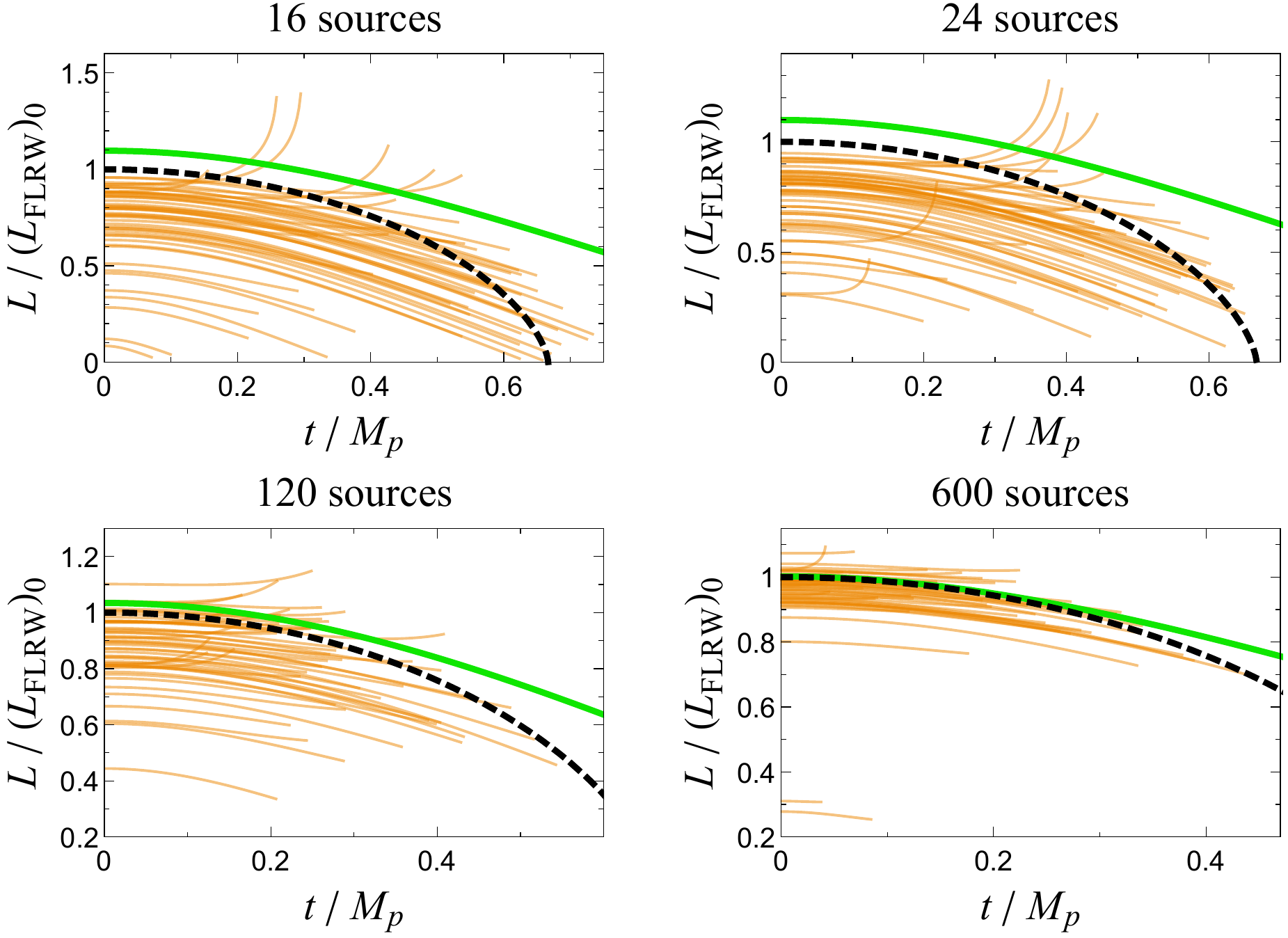}
\par\end{centering}
\caption{\label{fig: GCevolution 16}Circumference of LRS curves for configurations
consisting of 16 masses (orange). The green and black curve depicts
the regular model with 16 sources found in \cite{Clifton_etal:2012}
and the spherical matter dominated FLRW model respectively.}
\end{figure*}

A richer variety of dynamics can be seen for the remaining cases
shown in figures \ref{fig: GCevolution 16}. Looking
at them, we can identify two general qualitative categories. The first category consists 
of curves which decrease monotonically with time, such as the FLRW case and models with 
regular configurations. A majority of the DI models exhibits this behaviour. The second
category consists of curves which, at some point in time start to increase. 
This represents a turnaround in which the space in a neighbourhood of the LRS curve undergoes a change from contraction to expansion. 
This behaviour drastically deviates from the FLRW model
and the regular cases. We cannot say if this is representative of the entire universe or if it is a semi-local anomaly.

\subsection{Hubble and Deceleration Parameters}
There are two additional parameters we can measure which
enables further comparisons with FLRW cosmology. These are the Hubble
and deceleration parameters which we define as 
\begin{equation}
   \mathcal{H}_{L}:=\frac{\dot{L}}{L}
\end{equation}
\begin{equation}\label{eq: deceleration parameter definition} 
   q_{L}:=-\frac{\ddot{L}L}{\dot{L}^{2}}
\end{equation}
where $L$ is the length of the LRS curve. These two parameters are plotted for all the models we consider in figures \ref{fig_8_H_q}-\ref{fig:H_q_120_600}. The black dashed curve depicts as usual the spherical FLRW case.
\begin{figure}
\begin{centering}

\includegraphics[scale=0.7]{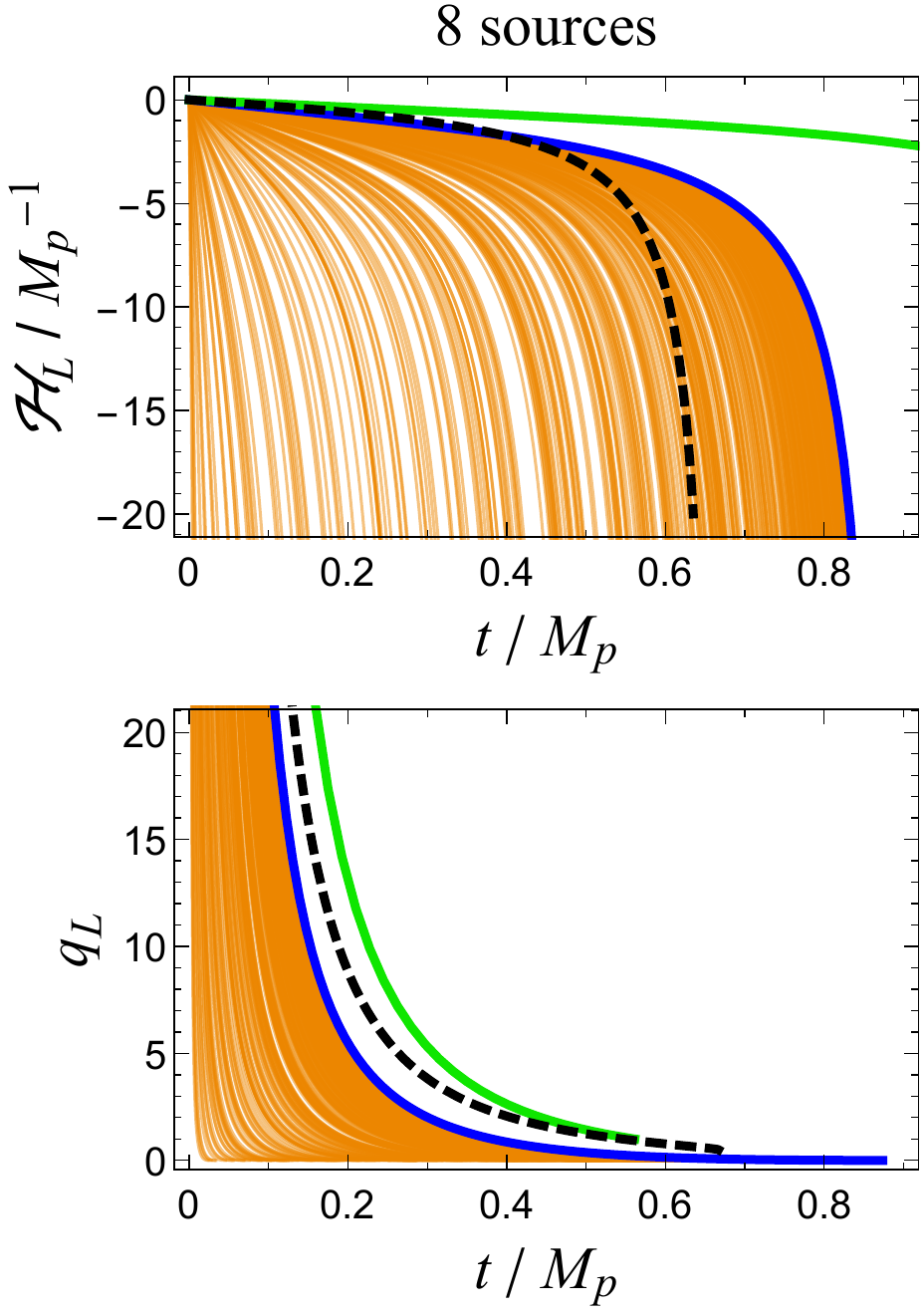}
\smallcaption{Figure depicting the Hubble parameter
and the deceleration parameter for 500 randomly generated
configurations consisting of 8 sources each. In the blue case, the source is located at the cell center. The green and dashed curves respectively
refer to the regular model with 8 sources given in \cite{Clifton_etal:2012} and the spherical matter dominated FLRW modely.}
\label{fig_8_H_q}

\par\end{centering}
\end{figure}
In the 8 source model (figure \ref{fig_8_H_q}), all curves show the same qualitative behaviour due to the position independent curvature. Even though this case is very different from the FLRW model in terms of size (see figure \ref{fig: LRS curve length plot}), it is qualitatively very similar.
\begin{figure*}
\begin{centering}
\includegraphics[scale=0.6]{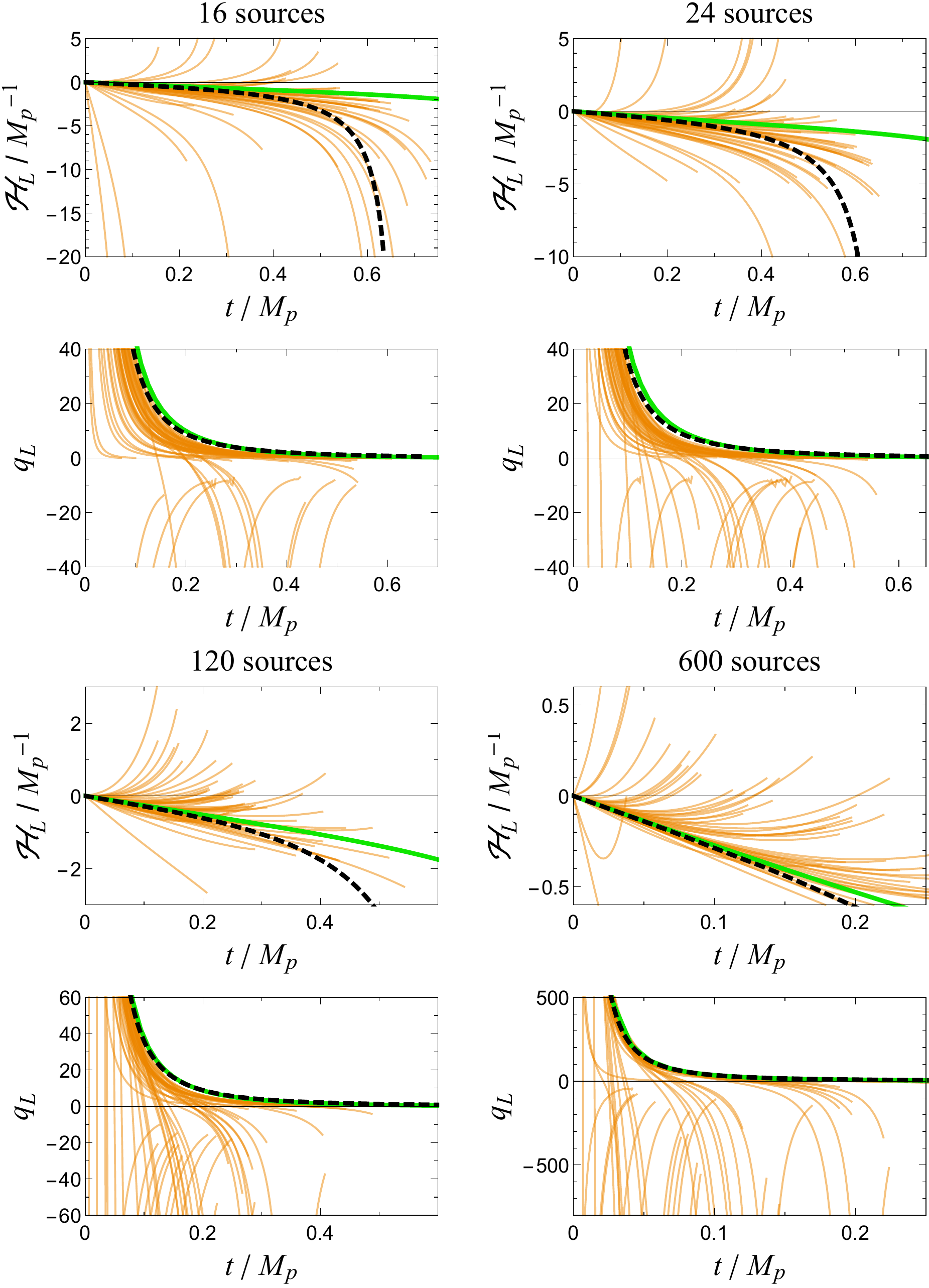}
\par\end{centering}
\caption{Figures depicting the Hubble parameter and the deceleration parameter for 50 randomly generated configurations consisting of 16 - 600 sources (orange). The green and black curves depict the regular models with corresponding number of sources, found in \cite{Clifton_etal:2012}, and the spherical matter dominated FLRW model respectively.}
\label{fig:H_q_120_600}

\end{figure*}


In the remaining models (figure \ref{fig:H_q_120_600}) the dynamics are more complex. Turning our attention to the deceleration parameter, we can broadly divide the behaviour in two
categories. First we have all the curves for which the deceleration
parameter is always positive. An example is provided in figure \ref{fig: deceleration examples}a. The curve starts at positive infinity before approaching zero. The FLRW curve and the regular models belong to this category, as well as a majority of the DI models.
%


\begin{figure*}
\begin{centering}
\includegraphics[scale=0.5]{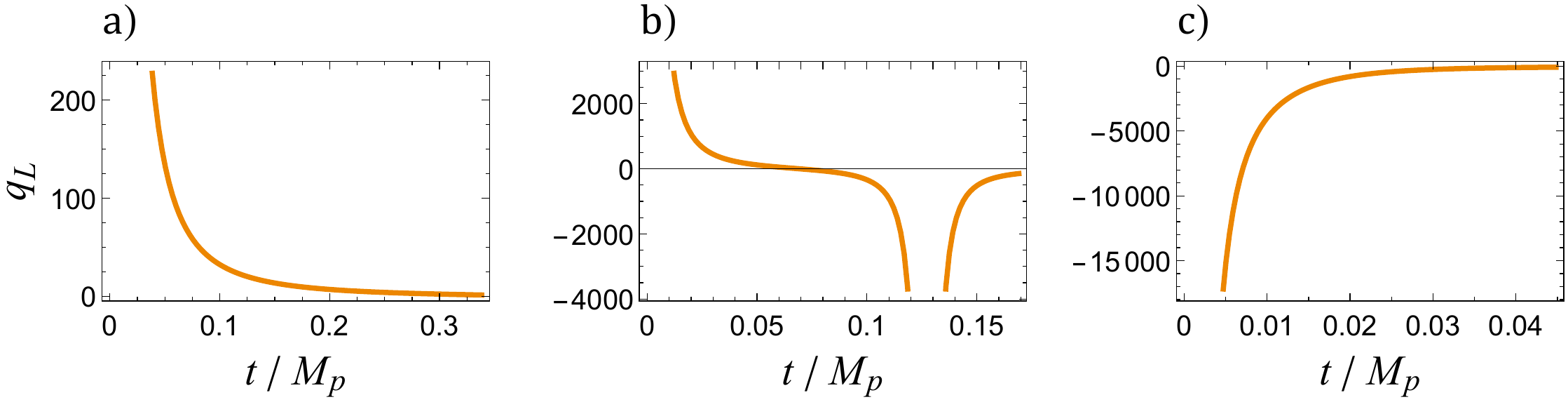}
\par\end{centering}
\caption{\label{fig: deceleration examples} Three possible behaviours of the deceleration parameter with 600 sources. Panel a) shows only positive values while b) switches sign at some later time. The singularity in b) is due to a vanishing first derivative (see equation (\ref{eq: deceleration parameter definition})). These are the most common types. We encounter occasionally deceleration parameters which are always negative, as depicted in c).}
\end{figure*}
The second category is characterized by the presence (at some point in time) of negative deceleration parameters. These curves are usually positive initially. An example is shown in figure \ref{fig: deceleration examples}b. In rare cases can we find models with always negative deceleration parameters, as in figure \ref{fig: deceleration examples}c. Regardless, the appearance of negative deceleration parameters is interesting, since in FLRW cosmology, negative values correspond to the presence of some exotic matter such as dark energy \cite{Goobar}. We show however that acceleration can occur across large regions of space in models with only ordinary mass (specifically Schwarzschild black holes) for some matter configurations.

Another interesting observation can be discerned from figures \ref{fig_8_H_q}-\ref{fig:H_q_120_600}. 
From the Hubble parameters we see that the fraction of orange curves which terminates below the
FLRW curve (black dashed curve) decreases as the source count 
increases. This provides a crude indication that the mean behaviour of the 
Hubble parameter becomes less Friedmann-like at late times as we increase the source count  
(if the mean was Friedmann-like, we would expect an equal number of curves to terminate 
above and below the FLRW case).

A valid question is whether or not our results would change
if we would have included the $\text{curl}H$ term. However, we know that the error starts from zero (being of order $\sim t^3$ initially), and as demonstrated in \cite{Bentivegna 2}
and to be illustrated in section \ref{sec: Calculating curlH},
the gravitomagnetic effect is very small.



\section{Fitting the Friedmann Equation}

The Friedmann equation for a late time flat cosmology consisting of dark energy and
ordinary matter can be expressed as \cite{Goobar}

\begin{equation}
\left(\frac{\dot{a}}{a}\right)^{2}=H_{0}^{2}\left(\Omega_{M_{0}}\left(\frac{a_{0}}{a}\right)^{3}+\Omega_{\Lambda_{0}}\right)\label{eq: Friedmann density}
\end{equation}
where $\Omega_{M_{0}}$ and  $\Omega_{\Lambda_{0}}$  are the
density parameters corresponding to mass  and dark energy respectively.
Furthermore the relationship between the density parameters are 

\begin{equation}
\Omega_{M_{0}}+\Omega_{\Lambda_{0}}=1.\label{eq: density conservation}
\end{equation}
By fitting expression \ref{eq: Friedmann density} to the blue curve in figure \ref{fig: curve fitting examples}, we obtain an indication of what kind of energy content is necessary for an approximate DI-like behaviour. We perform this fit on a single randomly generated DI model with 600 masses exhibiting generic dynamics in the LRS length evolution.

The quantities with a 0-subscript in equation (\ref{eq: Friedmann density})
indicate ``present day'' values, obtained through observation/measurement carried out 
at the ``present time'' $t_{0}$. We define $t_{0}$ to be the time when an observer residing in the DI universe carries out measurements or observations, indicated by the vertical dashed line in figure \ref{fig: curve fitting examples}.
The observer's time is considered to progress from right to left in the figure. Consequentially, the \emph{observer's future} is the 
initially time-symmetric hypersurface at $t=0$ and
the \emph{observer's past} is located at large $t$. The notions of future
and past are, with respect to the time axis in the figure, flipped
in order to represent an expanding universe. Only parts of the curve where 
$t>t_{0}$ (regions to the right of the dashed line) will be fitted against using equation (\ref{eq: Friedmann density}), 
since the observer possesses only information about the past.

By allowing the present time $t_{0}$ to progressively
increase, i.e. the observer will be moved further away (temporally) from
the initial time-symmetric hypersurface (located at $t=0$), there will be less and less portions of the simulated data available for our fit. This will alter
the density parameters obtained as a function of $t_0$. During the fitting, $\Omega_{M}$ is allowed to vary freely but is restricted to non-negative real numbers.

\begin{figure*}
\begin{centering}
\includegraphics[scale=0.55]{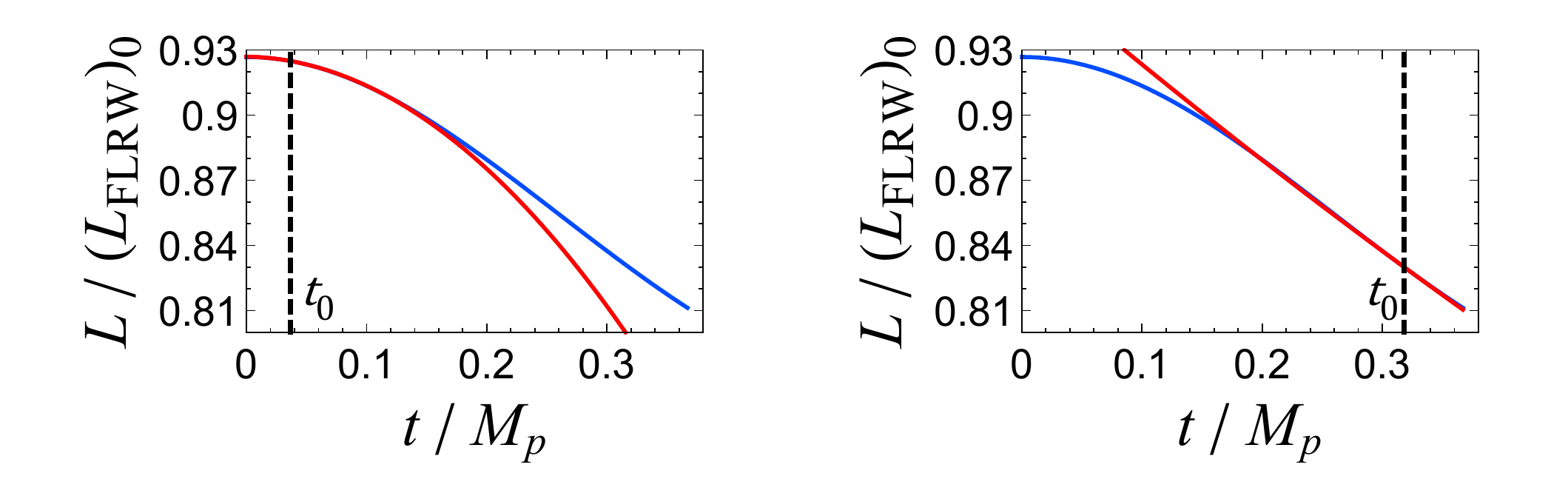}
\par\end{centering}

\caption{\label{fig: curve fitting examples} Examples of fitting equation (\ref{eq: Friedmann density}) (red)
to the LRS length in a randomly generated DI model (blue).
The ``present time'' $t_{0}$ is progressively increased, as is visible
by a shift of the vertical dashed line from left to right. The blue and red curves are
tangent to each other at $t_{0}$ since we require the fitted curve to have
the same $H_{0}$ and $a_{0}$ as our simulation.}
\end{figure*}

\begin{figure}
\begin{centering}
\includegraphics[scale=0.7]{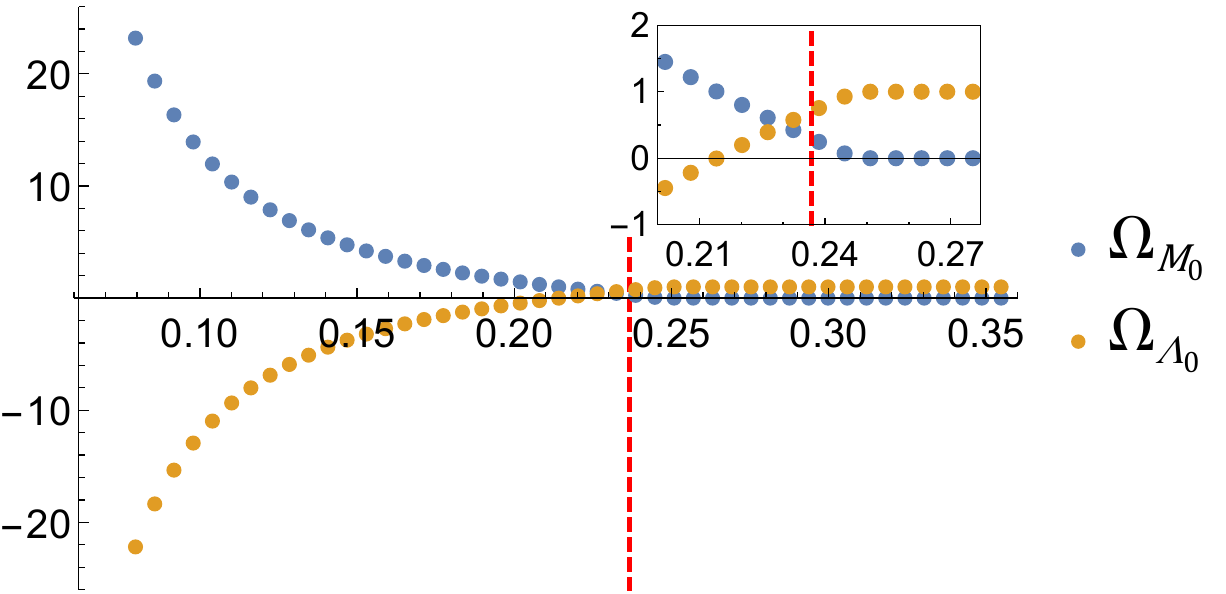}
\par\end{centering}

\caption{\label{fig: density parameters} The relationship between the fitted
density parameter $\Omega_{M_0}$  as a function
of time $t_0$. The value of $\Omega_{\Lambda_0}$ is derived using \eqref{eq: Friedmann density}. At the red dashed line we obtain values $\Omega_{M_{0}}=0.3$ and $\Omega_{\Lambda_{0}}=0.7$.}
\end{figure}

Figure \ref{fig: density parameters} depicts the density parameters
$\Omega_{M_{0}}$ and $\Omega_{\Lambda_{0}}$ as a function of 
$t_{0}$. For small $t_{0}$, i.e. close to the initial hypersurface,
we have $\Omega_{M_{0}}\gg1$ which therefore results in negative
$\Omega_{\Lambda_{0}}$. The inset is an enlargement on a region where both parameters are positive simultaneously. Within this region we can have that 
$\Omega_{M_{0}}=0.3$ (indicated by dashed red line), which corresponds well with current observations (see for instance  \cite{Planck satellite 2015}).


\section{\label{sec: Calculating curlH}Calculating the $\text{curl}H$-term} \label{calculating curlH}
In this section we will estimate the curl\emph{H} term and see how it evolves with time in order to determine the magnitude of the resulting error. The $\text{curl}H$-term 
in equation (\ref{dot_E}) can be evaluated recursively in
a truncated Taylor series over time. The first and second order terms
are zero, permitting us to stick with the third order term only, given by \cite{Clifton_etal:2017}
\begin{equation}
C_{+}^{(3)}=-n^{a}n^{b}\left[\text{curl curl}\left(E^{2}\right)_{ab}\right]_{t=0}t^{3}\label{eq: truncated curlH}
\end{equation}
where the tensor $\left(E^{2}\right)_{ab}$ is defined as
\begin{equation}
\left(E^{2}\right)_{ab}=E_{a}^{\,\,\,\, c}E_{cb}.
\end{equation}
Figure \ref{fig: ODE vs curlH for 24 masses} shows an example of
how the magnitude of the ODE expression (see equation (\ref{dot_E}))
compares to the curl\emph{H-}term (\ref{eq: truncated curlH}) as
a function of time for 24 masses on a random point on the LRS
curve. A general feature depicted in figure \ref{fig: ODE vs curlH for 24 masses}
is the initially zero curl\emph{H} term, which makes the differential
equation (\ref{dot_E}) valid for small \emph{t} .

\begin{figure}
\begin{centering}
\includegraphics[scale=0.6]{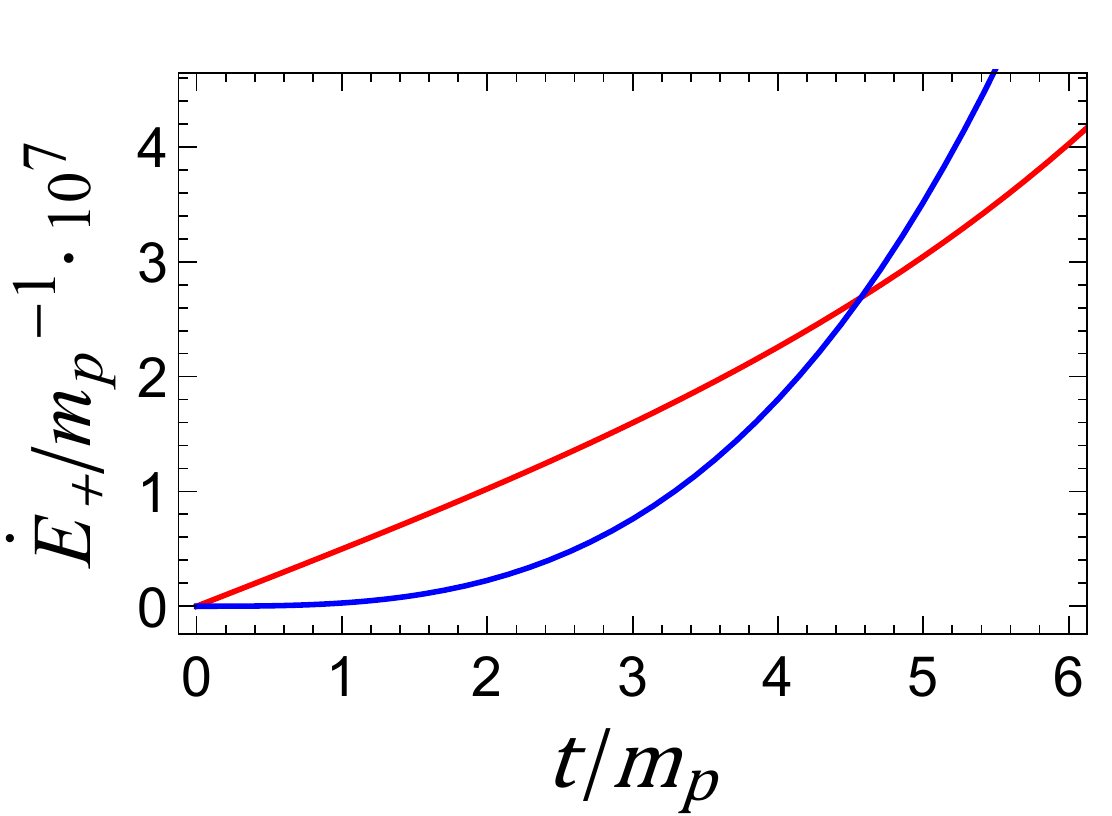}
\par\end{centering}

\caption{\label{fig: ODE vs curlH for 24 masses}The red curve depicts the
ODE \eqref{dot_E}, while the blue curve corresponds to the curl\emph{H} term  (\ref{eq: truncated curlH}). Notice
that the curl\emph{H} term is very small for early \emph{t}. Quantities
are given in terms of the median proper mass $m_{p}$.}
\end{figure}

We can make some statistics of when the ODE and curlH-term are equal
in magnitude. The result is shown in figure \ref{fig: curlH ODE intersection},
where we see the median time expressed in terms of the median proper
mass $m_{p}$. We managed to calculate expression (\ref{eq: truncated curlH})
for configurations consisting of 8, 16 and 24 masses only. For 120
and 600 masses, the evaluating $\left[\text{curl curl}\left(E^{2}\right)_{ab}\right]_{t=0}$
becomes a heavy computational endeavor and is therefore omitted. However
a clear pattern can be distinguished: as we increase the number of
black holes, the time it takes for the ODE and curl\emph{H} term to
become equal increases. If we compare figure \ref{fig: curlH ODE intersection} with figure \ref{fig: singularity statistics}, we notice that for 600 sources it appears as if coordinate singularities appear before the curl\emph{H} term becomes dominant.

\begin{figure}
\begin{centering}
\includegraphics[scale=0.6]{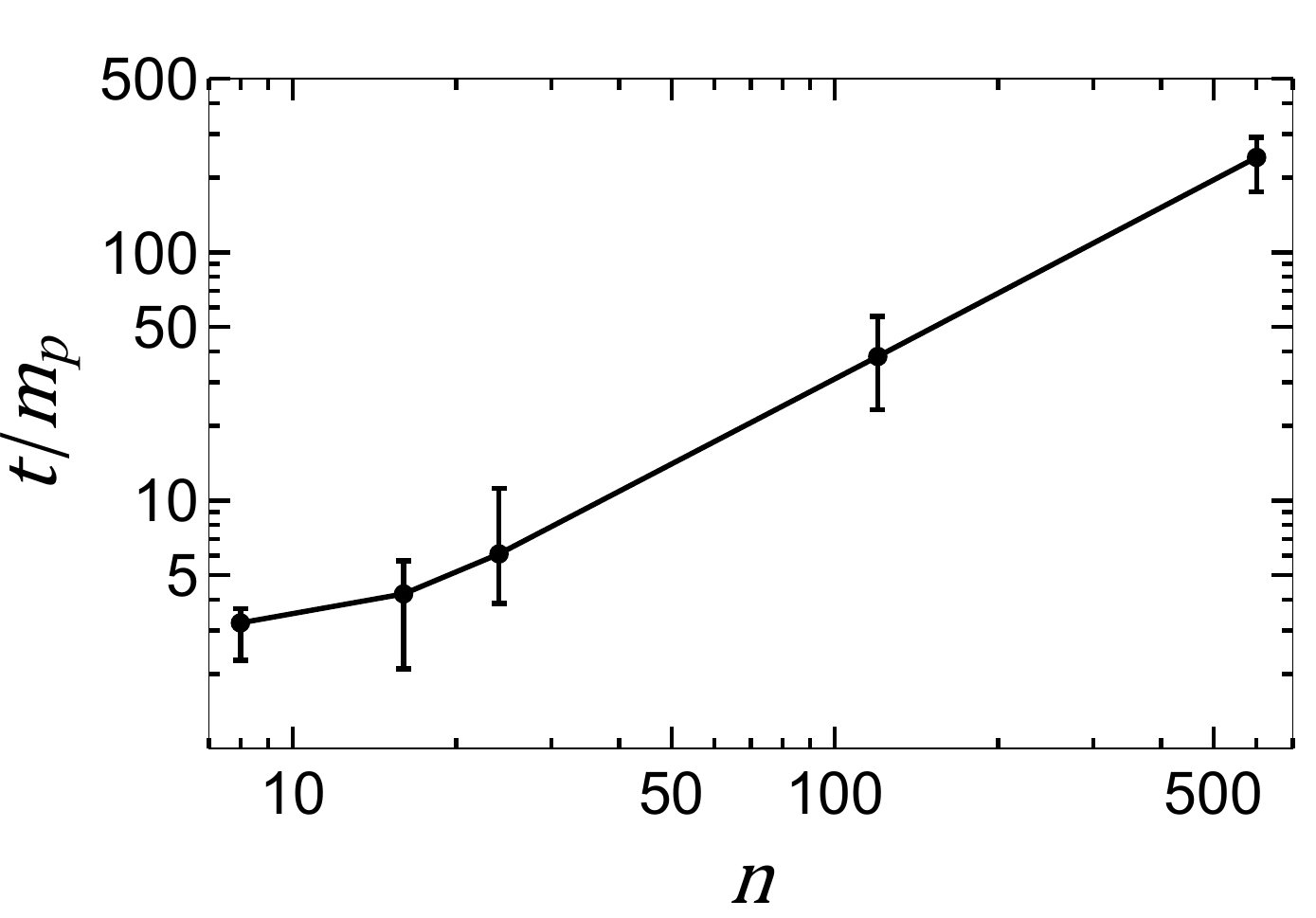}
\par\end{centering}

\caption{\label{fig: curlH ODE intersection}Plot depicting the median time
it takes for the curl\emph{H} term (\ref{eq: truncated curlH}) and ODE expression \eqref{dot_E} to become equally
large as a function of the number of sources \emph{n}. The error bars
indicate the quintiles.}
\end{figure}


\section{Evolution with Clusters of Black Holes}

\subsection{Generating Clusters}

In our universe, galaxies are not uniformly distributed but form instead
wide networks of filaments, clusters and voids. 
To simulate this behaviour to some extent, we have distributed black holes on the 3-sphere in such a way to encourage the formation of clusters.
The effect of discrete mass objects with some form of structure or clustering have already been investigated in for instance \cite{Korzynski_nonlinear_effects,Korzynski_backreaction_continuum_limit,Durk_Clifton}. 
Our approach however is stochastic and will use an algorithm inspired by the two-point function used
in galaxy surveys to quantify clustering \cite{Galaxy clusters}:
\begin{enumerate}
\item Focusing on one cell, distribute the desired number of black holes
in a random uniform way as described in subsection \ref{sub: uniform distribution of masses}.
\item Select one of the points randomly, \textbf{P}. The probability of
selecting any point is chosen to be equal. 
\item Measure the distance between \textbf{P} and every other point inside
the same cell. The probability of finding a mass at point \textbf{P}
is given by the expression
\begin{equation}
\text{Prob}\left(\mathbf{P}\right)=1-\prod_{i=1}^{n-1}\left(1-\left(\frac{d_{i}}{r_{0}}\right)^{-\gamma}\right),
\end{equation}
where $n$ is the number of masses, $d_{i}$ is the distance on the
3-sphere between \textbf{P} and another point \textbf{$\mathbf{P}_{i}$},
while $r_{0}$ and $\gamma$ are parameters. This gives the probability
of finding a mass at point \textbf{P }due to other masses located
at points \textbf{$\mathbf{P}_{i}$}.
\item Now perform a test. Generate a number, \textbf{T}, randomly and uniformly
from the interval $\left[0,\,1\right]$. If $\mathbf{T}\leq\text{Prob}\left(\mathbf{P}\right)$,
the test is passed. Leave point \textbf{P }alone and then proceed
to pick another point at random - start at step 2. If $\mathbf{T}>\text{Prob}\left(\mathbf{P}\right)$,
the test is failed. Move point \textbf{P }to another location. This
new location is determined randomly according to Marsaglia's algorithm
described in subsection \ref{sub: Marsaglia's algorithm}.
\item Using the new location, repeat step 3 and perform the test in step
4. Repeat steps 3 and 4 until the test is passed.
\item Once the test is passed, restart at step 2.
\end{enumerate}
This algorithm is repeated an arbitrary number of times. In our case,
we will only generate clusters in configurations containing 600 sources
and repeat the algorithm 6000 times. The parameters we used are $\gamma=3.5$
and $r_{0}=0.05$. We will henceforth refer to the DI model without any artificial clustering as \textit{uniform}.


\subsection{Dynamics for Configurations with Clusters}

Since clustering is only meaningful for many sources, we have exclusively
generated clusters in configurations with 600 black holes. The LRS circumference dynamics are plotted in figure \ref{fig: LRS length clustered and normal} in  purple for 50 random configurations. A general tendency one can notice is that the purple curves
are located further away from the homogenous limit given by the FLRW
model, than the orange curves. 
The median initial length for the uniform case is $0.96 (L_{FLRW})_0$ with a 60\% confidence interval ranging from $ 0.92 (L_{FLRW})_0$ to $ 0.99 (L_{FLRW})_0$.
This is closer to $1$ than the clustered case, with a median of $0.82 (L_{FLRW})_0$ and a confidence interval ranging from  $0.76 (L_{FLRW})_0$ to  $0.87 (L_{FLRW})_0$.
In addition, the clustered models have in general a longer "lifespan". 
The median time for the uniform model is $0.19 M_{p}$, but 74\% of the clustered models encounter a singularity later than this.  
A simple explanation is that clustered groups of black holes are treated as an effective discrete mass source. This causes a "grainier" mass distribution and moves the model away from the homogenous limit.

\begin{figure}
\begin{centering}
\includegraphics[scale=0.6]{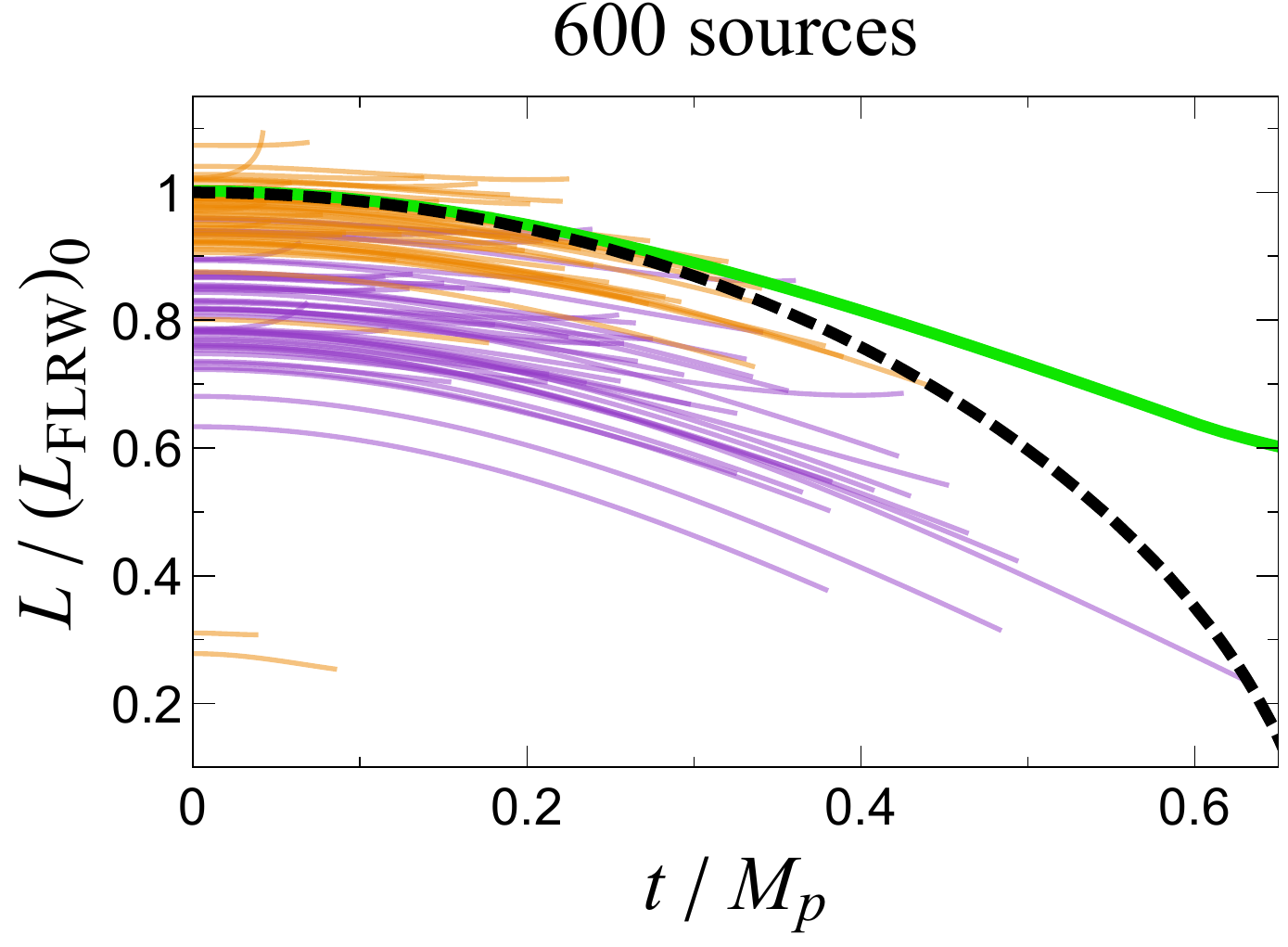}
\par\end{centering}

\caption{\label{fig: LRS length clustered and normal}The LRS circumference
and their evolution for four different models with 600 black holes.
The purple curves correspond to the DI model with clustered configurations.
The orange curves correspond to the uniform DI model with 600 sources.
The black dashed curve and the green curve correspond respectively
to the matter dominated spherical FLRW model and the regular model
with 600 sources found in \cite{Clifton_etal:2012}.}

\end{figure}

Figure \ref{fig: clustered hubble} plots the Hubble and deceleration parameters.
Turning our attention to the Hubble parameters, 60\% of the clustered configurations in purple lie above the FLRW curve at $t=0.2 M_{p}$, compared to 98\% of the
uniform configurations in orange. 
Comparing with figure \ref{fig:H_q_120_600}, decreasing the number of sources is accompanied by a decrease in the percentage of curves lying above the FLRW value at  $t=0.2 M_{p}$ , from the aforementioned 60\% down to 12\% for 8 sources.  
Thus such behaviour is more closely associated with configurations containing fewer sources than 600. 
This further strengthens our interpretation that a group of clustered black holes are considered as a single effective discrete mass.

\begin{figure}
\begin{centering}
\includegraphics[scale=0.7]{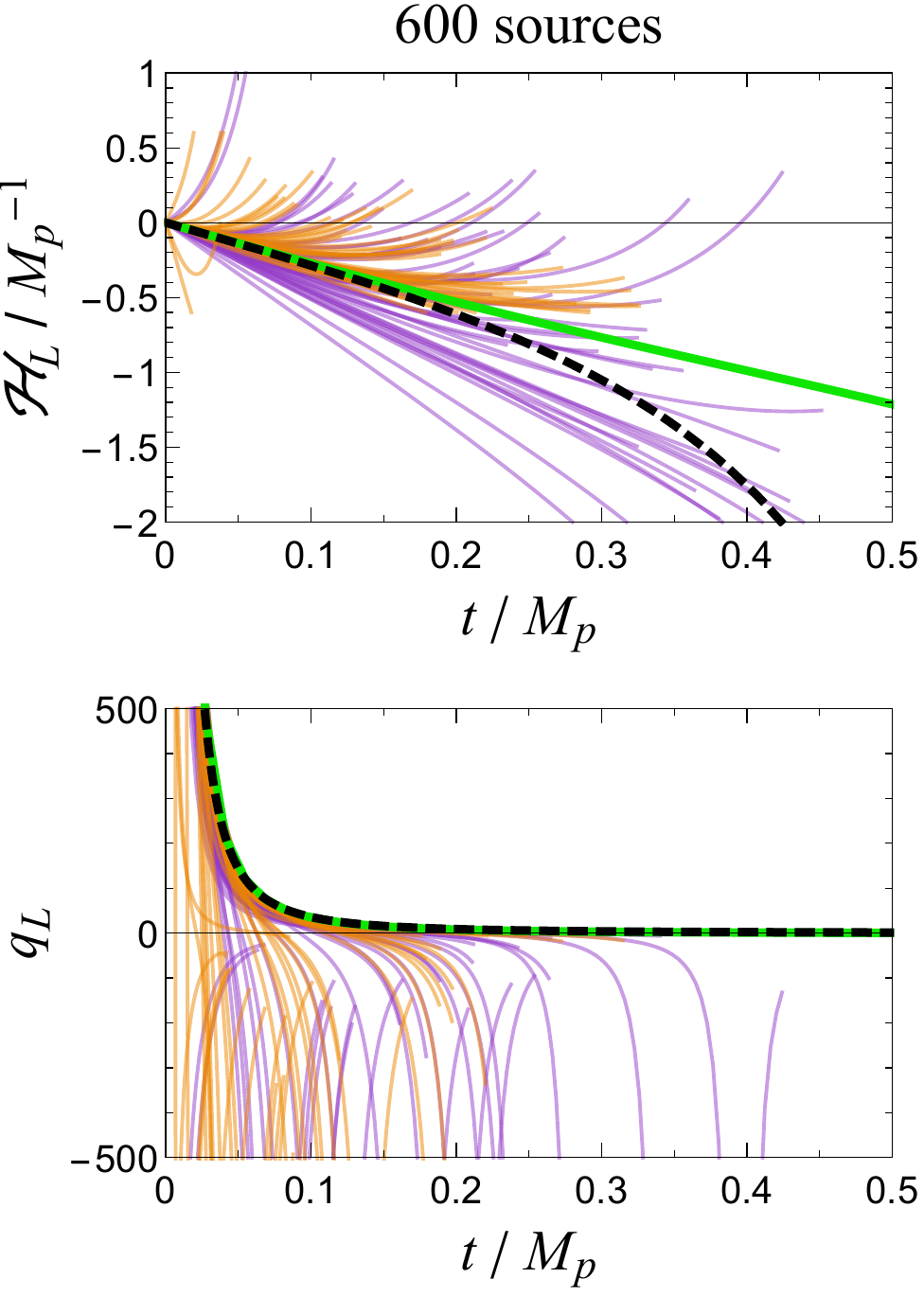}
\par\end{centering}

\caption{\label{fig: clustered hubble}The above and bottom plots depict respectively the evolution of the
Hubble parameter and deceleration parameter for four different models
with 600 sources. The purple curves correspond to the DI model with
clustered configurations while the orange curves correspond to the
uniform DI model. The black dashed curve and the green curve
correspond respectively to the matter dominated spherical FLRW model
and the regular model with 600 sources found in \cite{Clifton_etal:2012}.}
\end{figure}


\section{Concluding Remarks }

In earlier work on cosmological models with discrete sources, the focus has been on source distributions adapted to regular tessellations of spatial hypersurfaces. In such models, the sources reflect the regular nature of the models by being positioned at lattice centers. The LRS curves present in those models allow for a simplification of their dynamics due to the LRS symmetry.
In this paper we have presented the DI cosmological model consisting of discrete sources in the form of Schwarzschild black holes having an irregular distribution. To maintain the possibility to do simplified dynamics, we have used a special tessellation of $S^3$ in which there is a single LRS curve. This work therefore represents a generalization of previously discussed discrete models by allowing sources to be distributed irregularly while retaining the possibility of analytic analysis of the dynamical evolution. 

Our simulations suggest that initial conditions, such as the relation between total mass and size of the universe, approach the FLRW limit as the number of masses increases (see 
figure \ref{fig: LRS curve length plot}). However, the approach is slower than that of the regular models. Also, the approach in our models is from below rather than from above in the regular case. This is possibly related to the different types of tessellations used. When it comes to the dynamics, the DI models, as expected, exhibit a richer variety of behaviour than that seen in models with a regular configuration of mass sources. In fact, we find that the system is quite sensitive to the initial mass configuration. Other interesting behaviour is the appearance of negative  deceleration parameters.

The DI model has proven to be a simple and flexible model which permits the investigation of a large number of discrete sources distributed in various configurations. By using DI models with a larger number of sources one could determine more precisely the relationship between initial mass configuration and the resulting dynamics. It may also be possible to improve the DI model by using geodesic slicing in order to avoid coordinate singularities. One question which is raised by the DI model is whether the expanding acceleration sometimes observed is shared by other regions beside the LRS curve. Occurrence of large scale acceleration could be due to structure formation, often referred to as cosmic backreaction, as discussed by several authors (see for example \cite{Buchert&Rasanen:2012}). Our results for the DI models with built-in clustering show that their dynamics generally deviates more from that of the FLRW model compared to models without clustering even though the particular results for acceleration are less clear. 
Illuminating these issues could improve our understanding of how mass inhomogeneities affect the large scale geometry and dynamics of the universe. 

One obvious drawback of our DI model is that it has a certain amount of intrinsic anisotropy. This is due to the presence of a preferred great circle as part of the construction leading to a preferred global direction. 
This drawback can actully be overcome by noting that besides the DI model used here it is also possible to construct DI models based on a regular lattice by the same method. To preserve the LRS property in that case when performing reflections with irregular configurations, it is necessary that the number of cells meeting at an edge is an even number. The only regular model which fulfills this requirement is the 16-cell. However, all the regular models could actually be used for this purpose if one relaxes the regularity slightly. This can be done by noting that each 3-cell in a regular tessellation is intersected by a number of symmetry planes. These planes divide the 3-cell in subcells called chambers \cite{Borovik&Borovik:2010}. Any two chambers in a given regular tessellation are either identical or mirror images of each other. Since adjacent chambers are related by a reflection, one can pick one chamber as the seed cell and then distribute its configuration by reflection to all the other chambers. By such a construction based on the chambers instead of the regular cells themselves, the total number of sources corresponding to a given number of sources in a seed cell can be increased significantly. For example, a regular tetrahedron has 24 chambers and so a 600-cell has $24 \times 600 = 14400$ chambers.


\paragraph{Acknowledgements }
The authors wish to thank M. Korzy\'{n}ski for helpful comments and a referee for constructive remarks.


\end{document}